\documentclass[review,onefignum,onetabnum]{siamart190516}

\usepackage{amsfonts}
\usepackage{multirow}
\usepackage{graphicx}
\usepackage{epstopdf}
\usepackage{algorithmic}
\ifpdf
  \DeclareGraphicsExtensions{.eps,.pdf,.png,.jpg}
\else
  \DeclareGraphicsExtensions{.eps}
\fi

\newcommand{\realpart}{\mbox{$\mathrm{Re}$}}

\newcommand{\rank}{\mbox{${\mathrm{rank}}$}}

\newsiamthm{problem}{Problem}
{\theorembodyfont{\rmfamily} \newtheorem{example}[theorem]{Example}}

\newsiamthm{claim}{Claim}
\newsiamremark{hypothesis}{Hypothesis}
\crefname{hypothesis}{Hypothesis}{Hypotheses}
\newsiamremark{remark}{Remark}

\headers{Synchronization of Coupled Phase Oscillators}{Kaihua Xi and Zhen Wang {et al}}

\title{
  Synchronization of Coupled Phase Oscillators \\with Stochastic Disturbances and \\the Cycle Space of the Graph
}

\author{
Kaihua Xi
  \thanks{School of Mathematics, Shandong University, Jinan, 250100, 
    Shandong Province, China%
    (
    \email{kxi@sdu.edu.cn, wangzhen2017@mail.sdu.edu.cn, aijie@sdu.edu.cn}%
    ).%
  }
\and
Zhen Wang 
  \footnotemark[1]
\and
Aijie Cheng
  \footnotemark[1]
\and
Hai Xiang Lin
  \thanks{Delft Institute of Applied Mathematics, Delft University of Technology,
    Delft,2628CD,The Netherlands%
    (\email{H.X.Lin@tudelft.nl},
    \email{J.H.vanSchuppen@tudelft.nl}%
    ).%
  }
  \and 
  \newline
  ~~~~~~~~~~Jan H. van Schuppen
\footnotemark[2]
 \and
 Chenghui Zhang
   \thanks{School of Control Science and Engineering, Shandong University, Jinan, 250061,
    Shandong Province, China%
    (\email{zchui@sdu.edu.cn}%
    ).%
  }
  }
\usepackage{amsopn}

\begin{document}
\newsiamthm{assumption}{Assumption}
\maketitle
\begin{abstract}
The synchronization stability of a complex network system of coupled phase oscillators is discussed.   
In case the network is affected by disturbances,
a stochastic linearized system of the coupled phase oscillators
may be used 
to determine the fluctuations of phase differences 
in the lines between the nodes and 
to identify the vulnerable lines that may lead to desynchronization. 
The main result is the derivation of the asymptotic variance matrix 
of the phase differences
which characterizes the severity of the fluctuations.  It is found that the cycle space of the graph of the system plays a role in this
characterization. With theory of the cycle space, the effect of forming small cycles on the fluctuations are 
evaluated. It is proven
that adding a new line or increasing the coupling strength of a line affect the fluctuations in the lines in any cycle including this line while it does not affect the fluctuations in the other lines. In particular, if the phase 
differences at the synchronous state are not changed by these actions, then the affected fluctuations reduce. 
\end{abstract}
\section{Introduction}
\label{sec:intro}
Synchronization in a networked system of coupled phase oscillators
serves as a paradigm 
for understanding of the collective behavior 
of a real complex networked system. A coupled oscillator network is characterized by a population of heterogeneous oscillators and a graph describing the interaction among the oscillators.
Examples of such systems arise in nature 
(e.g., Kuramoto oscillators \cite{Kuramotomodel}, chimera spatiotemporal patterns \cite{Abrams}, 
cardiac pacemaker cells \cite{cardiac}) and in technological systems 
(for example, multi-agent systems \cite{macuiqin}, consensus problems \cite{Dorfler2012}, 
distributed optimization \cite{YANG2019278}, 
power grids \cite{Xi2016,motter}).  

\par
In this paper, we focus on systems which need synchronization for proper functioning, such as power
grids. If the synchronization is lost, then the systems can no longer function properly.  
The vast literature shows that
significant insights have been obtained on the emergence of a synchronous state 
(defined to be a steady state of the system) and
synchronization coherence. The synchronization 
is determined by the system parameters, including the natural frequencies at the nodes, the network
topology and the coupling strength of lines.  
With the metrics of the critical coupling strength \cite{Dorfler2012,Dorfler20141539,FAZLYAB2017181} and the 
order parameter \cite{Skardal2014}, the effects of these parameters on the synchronization are widely investigated. Based on these investigations, the system parameters may be assigned to optimize the synchrony, which can be attained 
by deletion or addition of lines, or 
by changing the coupling strength of the lines in the network. 
An important problem is to maintain the synchronization when the system is subjected to disturbances. 
Regarding the ability to maintain the synchronization, the spectrum of the system matrix of the linearized system and the volume of the basin of attraction of a stable synchronous state may be investigated. However, 
in these investigations, the severity of the disturbances are not considered and the lines at which the synchronization may be lost cannot be effectively identified.  The question how disturbances spread through networks of power systems, has attracted wide interest of investigation with  a toolbox for simulations and with analytic methods \cite{Delocalization,topological-spreading,Aue17,Networksuseptbilities}.
In control theory, the synchronous state 
is mentioned as \emph{the set point} for control, in which control actions are taken
to let the state converge to the synchronous state after disturbances.
Thus, under continuous disturbances, the
phase may fluctuate around the synchronous state. If the fluctuations of the phase differences are larger than the threshold $\pi/2$, a synchronous state may not be attained any longer, and then the synchronization may be lost \cite{Kuramotostability}.  This indicates 
the necessity to study the phase difference in the lines but not the phases at the nodes. One says that a line is {\em vulnerable} if the desynchronization occurs at this line easily.  Clearly, the lines with large fluctuations in the phase difference are
vulnerable. 
Modelling the disturbances by inputs to the system, the $\mathcal{H}_2$ norm of a linear input-output system has 
been widely used to study the fluctuations of the phase differences \cite{optimal_inertia_placement,H2norm}. 
With this $\mathcal{H}_2$ norm, the fluctuations may be effectively suppressed by assigning 
the system parameters, thus improve the robustness of the system.  
However, because the $\mathcal{H}_2$ norm equals to the trace of a matrix \cite{H2norm}, which 
is a global metric characterizing the sum of the fluctuations, the fluctuations of the phase differences 
in the lines and their correlation can hardly be explicitly characterized.

\par 
In this paper, we investigate the dependence of the fluctuations  of the phase differences in each line on the system parameters analytically, which can be used to suppress the fluctuations and identify the vulnerable lines, thus improve 
the ability of the system to maintain the synchronization. We model the disturbance by a set of Brownian motions
and reformulate the system as a stochastic linear system.
It is well known that for a linear Gaussian stochastic system with a system matrix that is Hurwitz, there exists an invariant probability distribution of the state that is a Gaussian probability distribution characterized by the mean value and the asymptotic variance of the state \cite[Theorem 1.53]{linearOptimalSystems}\cite[Theorem 6.7]{karatzas:shreve:1988}.
In the invariant distribution, the asymptotic variance characterizes the severity of the fluctuations in the phase difference in each line
of the system. The focus of this paper
is on the asymptotic variance
of a stochastic linearized network system of coupled phase oscillators, which reveals how the fluctuations in the phase differences depend on the system parameters. With the asymptotic matrix as a metric, a new avenue is open to study the robustness of network systems against 
the disturbances.
The contribution of this paper include explicit formulas of the asymptotic variance matrix and findings 
from these formulas on the impact of adding new lines and strengthening the coupling of the oscillators. 
To the best knowledge of the authors, for the first time the cycle space of a graph is related to the robustness of the system by an
explicit formula.
The method to study the synchronization stability in this paper can be extended 
to the networks with synchronizations, such as the power systems \cite{ZWang} and the general diffusive network 
\cite{Skardal2014}.
\par
The paper is structured as follows.
Section~\ref{sec:kuramotosystem}
provides elementary preliminaries on graphs theory and the invariant probability 
distribution of stochastic process. 
We 
formulate problem of the complex network of coupled phase oscillators 
and present the main results on the asymptotic variance in Section~\ref{sec:asymptoticvariance}.
The findings from the asymptotic variance are illustrated in three example networks in Section~\ref{sec:examples}.
Section~\ref{sec:proofs} provides the proofs of the results and Section \ref{sec:conclusions} concludes with remarks.
%
%
%
%
\section{Preliminaries}\label{sec:kuramotosystem}
The elementary notation, properties of graphs and the cycle space and the concept of 
the asymptotic variance of a stochastic Gaussian system are introduced in this section. 
\subsection{Notations}
\par
The set of the integers is denoted by
$\mathbb{Z} = \{ \ldots, ~ -1, ~ 0, ~ 1, ~ 2, \ldots \}$
and that of the positive integers by
$\mathbb{Z}_+ = \{ 1, ~ 2, ~ \ldots \}$.
For any integer $n \in \mathbb{Z}$
denote the set of the first $n$ positive integers by
$\mathbb{Z}_n = \{ 1, ~ 2, ~ \ldots, ~ n \}$.
The set of the real numbers is denote by $\mathbb{R}$.
Denote the strictly positive real numbers by $\mathbb{R}_{+} = (0, ~ +\infty)$.
\par
The vector space of $n$-tuples of the real numbers
is denoted by $\mathbb{R}^n$ for an integer $n \in \mathbb{Z}_+$.
For the integers $n, ~ m \in \mathbb{Z}_+$
the set of $n$ by $m$ matrices with entries of the real numbers,
is denoted by $\mathbb{R}^{n \times m}$.
Denote the identity matrix of size $n$ by $n$ by
$\mathbf{I}_n  \in \mathbb{R}^{n \times n}$,
which may also be denoted by $\mathbf{I}$ 
if the size is clear from the context.
\par
Denote subsets of matrices according to:
for an integer $n \in \mathbb{Z}_+$,
$\mathbb{R}_{spd}^{n \times n}$ denotes
the subset of symmetric positive semi-definite matrices
of which an element is denoted by $0 \preceq \mathbf{Q} = \mathbf{Q}^{\top}$; 

for matrices $\mathbf{A}\in\mathbb{R}^{n\times n}$ and $\mathbf{B}\in\mathbb{R}^{n\times n}$, denote by
$\mathbf{A}\preceq \mathbf{B}$ that $\mathbf{B}-\mathbf{A}$ is semi-positive-definite;
$\mathbb{R}_{nsng}^{n \times n}$
the subset of nonsingular square matrices;
$\mathbb{R}_{ortg}^{n \times n}$
the subset of orthogonal matrices
which by definition satisfy 
$\mathbf{U} ~ \mathbf{U}^{\top} = \mathbf{I}_n = \mathbf{U}^{\top} ~ \mathbf{U}$.
Call a square matrix $\mathbf{A} \in \mathbb{R}^{n \times n}$ {\em Hurwitz}
if all eigenvalues have a real part which is strictly negative;
in terms of notation,
for any  eigenvalue 
$\lambda(\mathbf{A})$ of the matrix $\mathbf{A}$, 
$\realpart (\lambda(\mathbf{A})) < 0$.
\par 
\subsection{Graphs and the Cycle Space}\label{subsection:graph}
Consider an undirected weighted network $\mathcal{G}=(\mathcal{V},\mathcal{E})$
with a set of $n \in \mathbb{Z}_+$ nodes denoted by $\mathcal{V}$ and 
a set of $m \in \mathbb{Z}_+$ edges or lines denoted by $\mathcal{E}$ and line weight $w_{ij}=w_{ji}\in\mathbb{R}_{+}$ if the nodes $i$ and $j$ are connected  and $w_{ij}=0$ otherwise. Denote by $k = (i, ~ j) \in \mathcal{E}$
the edge connecting the nodes $i$ and $j$
which edge is also denoted by $k$.
The Laplacian 
matrix of the graph is defined as  $\mathbf{L}=(l_{{ij}})\in\mathbb{R}^{n\times n}$ with
 \begin{eqnarray}\label{LaplacianDef}         
          l_{ij} =
          \left\{
          \begin{array}{ll}
            - w_{ij},                               & \mbox{if} ~ i\neq j,\\
            - \sum_{k=1, ~ k \neq i}^n ~ w_{ik}& \mbox{if} ~ i=j.
          \end{array}  
          \right. \nonumber 
\end{eqnarray}
The incidence matrix is defined as $ \mathbf{C} =(C_{ik}) \in \mathbb{R}^{n\times m}$ with $C_{ik}\in\mathbb{R}$,
\begin{eqnarray}\label{IncidenceDef}  
        C_{ik} 
    & = & \left\{
          \begin{array}{rl}
             1, & \text{if node $i$ is the beginning of line $e_k$},\\
            -1, & \text{if node $i$ is the end of line $e_k$},\\
             0, & \text{otherwise},
          \end{array}  
          \right.
          \end{eqnarray}
 Here the direction of line $e_k$ is arbitrarily specified in order to define the incidence matrix. 
 We further define a diagonal matrix $\mathbf{R}= \text{diag} (R_k)\in\mathbb{R}^{m\times m}$ with $R_{k} = w_{ij}$
 being the weight of line $e_k$ with $k=(i,j)\in\mathcal{E}$.  
 Elementary properties of matrices which are needed subsequently are summarized in 
 the next lemma.
 \begin{lemma}\label{lemma:laplacianmatrixproperties}
Consider the graph $\mathcal{G}$ with matrices $\mathbf{L}, \mathbf{C}, \mathbf{R}$.
\begin{itemize}
\item[(i)] The Laplacian matrix $\mathbf{L}$ is symmetric and hence all its eigenvalues are real.
\item [(ii)] Following the Gerschgorin' theorem  \cite[Theorem 36]{PietVanMieghem2008}, all the eigenvalues of $\mathbf{L}$ are non-negative. 
\item[(iii)] Denote the eigenvalues of $\mathbf{L}$ by $0\leq\mu_1\leq\mu_2\leq\cdots\leq \mu_n$. 
It holds
$\mathbf{L}  \mathbf{1}_n = \mathbf{0}_n$, thus, $\mu_1=0$ is an eigenvalue of $\mathbf{L}$ with an eigenvector $\tau \mathbf{1}_n$ where $\tau\in\mathbb{R}$.
\item [(iv)] The graph $\mathcal{G}$ is connected if and only if the second smallest eigenvalue $\mu_2>0$ \cite[Theorem 10]{PietVanMieghem2008}.  
\item [(v)] A relation between the incidence matrix $\mathbf{C}$ and 
the Laplacian matrix $\mathbf{L}$ is
\begin{eqnarray}\label{incidenceLap}
\mathbf{C} \mathbf{R} \mathbf{C}^{\top} = \mathbf{L}.
\end{eqnarray}
\item[(vi)]
It holds
$\mathbf{C}^{\top} \mathbf{1}_n = \mathbf{0}_m$.
\item [(vii)]
If the graph $\mathcal{G}$ is connected, then $rank(\mathbf{C})=n-1$ \cite[Theorem 1]{PietVanMieghem2008}. 
 \end{itemize}
 
\end{lemma}

\par 
The concept of the cycle space of the graph
plays an important role in the characterization
of the asymptotic variance matrix in this paper, which is defined below.
\begin{definition}\label{def:cyclespaceofgraph}
Consider a connected and undirected graph $\mathcal{G}=(\mathcal{V},\mathcal{E})$ with matrix $\mathbf{C}$.
\begin{itemize}
\item[(i)] 
If $\mathcal{C}$ is a subset of $\mathcal{E}$ such that the subgraph formed by $\mathcal{C}$ is a cycle graph,in which there are at least three nodes, the number of nodes equals to the number of lines and all the nodes are in 
a path that starts and ends at the same node without repeating any lines, then 
$\mathcal{C}$ is a cycle in $\mathcal{G}$.
\item[(ii)] Let $\mathcal{T}$ be a spanning tree of $\mathcal{G}$,
then there are $n-1$ edges in $\mathcal{T}$ and $m-n+1$ edges of $\mathcal{G}$ lie outside 
of $\mathcal{T}$. For each of these $m-n+1$ edges $e\in\mathcal{E}\backslash\mathcal{E}(\mathcal{T})$, the graph $\mathcal{T}+e$ contains a unique cycle in which the lines forms a fundamental cycle of the graph $\mathcal{G}$.
\item[(iii)] The {\em cycle space} of graph $\mathcal{G}$
is defined as the kernel of the incidence matrix $\mathbf{C}$,
\begin{eqnarray*}
        \mathbf{X}_{cysp}
    & = & \ker (\mathbf{C})
          = \{ \mathbf{\xi} \in \mathbb{R}^m |~ 
               \mathbf{C} \mathbf{\xi} = \mathbf{0}_n \} \subseteq \mathbb{R}^m, \\
        \dim(\mathbf{X}_{cysp}) 
    & = & n_{cysp} 
          = m - \rank(\mathbf{C})
          = m - n + 1.
\end{eqnarray*}
\end{itemize}
\end{definition}
\par 
The basis vectors of the cycle space can be derived based on the following theorem.
\begin{theorem}\label{theoremcyclespace}
\cite[Theorem 4.5,Theorem 5.2]{NORMAN}\cite[Chapter 4]{Graphtheory}
Consider a connected and undirected graph $\mathcal{G}=(\mathcal{V},\mathcal{E})$ with matrix $\mathbf{C}$.
\begin{itemize}
\item[(i)]For a cycle $\mathcal{C}_c$ 
with set $\mathcal{E}_{c}$ of lines in the graph $\mathcal{G}$, 
we specify a direction for $\mathcal{C}_c$ and define a vector for the cycle, 
\begin{eqnarray*}
        \mathbf{\xi}_c 
    & = & \begin{bmatrix}
     \xi_{c,1} & \xi_{c,2} & \cdots & \xi_{c,m}
    \end{bmatrix}^\top\in \mathbb{R}^{m},\\
        \xi_{c,k}
    & = & \begin{cases}
           +1, & \text{ if $e_k\in\mathcal{E}_c$ 
                        with direction $=$ the cycle direction
                      },\\
           -1, & \text{ if $e_k\in\mathcal{E}_c$ 
                        with direction $\neq$ the cycle direction
                      },\\
            0, & \text{ otherwise,}
        \end{cases}
\end{eqnarray*}
Then, it satisfies $\mathbf{C} \mathbf{\xi}_c = \mathbf{0}$, thus, $\mathbf{\xi}_c$ belongs to the kernel of $\mathbf{C}$. 
\item[(ii)] A {\em set of basis vectors} of the cycle space can be derived 
by taking the vectors as $\mathbf{\xi}_c$ for $c=1,\cdots,m-n+1$
corresponding to the $(m-n+1)$~ \emph{fundamental cycles} 
\end{itemize}

\end{theorem}
\par 
If the direction of the cycle $\mathcal{C}_c$ is changed, the vector $-\xi_c$ is obtained, which can also be a basis vector of the kernel of the cycle space. Thus, to obtain the basis vectors of the cycle space, the directions of the cycles can be specified arbitrarily, which is independent of the directions of the lines specified for the definition of the incidence matrix. 
\par
\begin{figure}[t]
\centering
\includegraphics[scale=1.0]{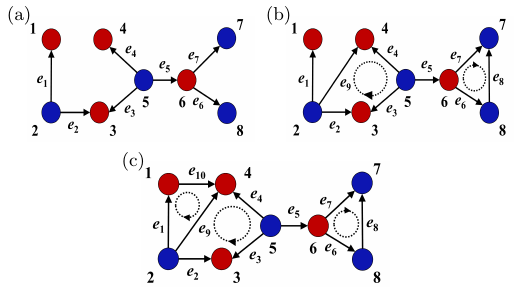}
\caption{Three networks with 8 nodes.}\label{fig1}
\end{figure}

 
We take the networks in Fig.\ref{fig1} as examples to illustrate 
the cycle space of graphs and its basis vectors. The directions of the lines are arbitrarily specified, and 
the directions of all the cycles are chosen to be clockwise. These directions are set for 
the calculation of the incidence matrix $\mathbf{C}$ and the basis vectors of the cycle space, 
which does not mean the networks are directed. 
The network (a) is a tree network hence $m=n-1$, thus the dimension of the kernel of the incidence 
matrix is zero and no basis vectors can be formulated. For network (b), the basis vectors
of the cycle space are $\mathbf{\xi}_1=[0,0,0,0,0,-1,1,-1,0]^{\top}$ and $\mathbf{\xi}_2=[0,-1,1,-1,0,0,0,0,1]^{\top}$ corresponding to the cycles $\{e_2,e_3,e_4,e_9\}$
and $\{e_6,e_7,e_8\}$ respectively. For network (c), the basis vectors of the cycle space 
are  $\mathbf{\xi}_1=[0,0,0,0,0,-1,1,-1,0]^{\top}$, $\mathbf{\xi}_2=[0,-1,1,-1,0,0,0,0,1]^{\top}$ and $\mathbf{\xi}_3=[1,0,0,0,0,0,0,0,-1,1]$ corresponding to the cycles $\{e_2,e_3,e_4,e_9\}$
and $\{e_6,e_7,e_8\}$ and $\{e_1,e_9,e_{10}\}$ respectively.
 Note 
that for the graph with cycles, because the spanning tree may not be unique, the set of the basis vectors of the cycle space may be non-unique. For example, for network (c), the basis vectors of the cycle space can also be 
$\mathbf{\xi}_1=[0,0,0,0,0,-1,1,-1,0]^{\top}$, $\mathbf{\xi}_2=[0,-1,1,-1,0,0,0,0,1]^{\top}$ and $\mathbf{\xi}_3=[1,-1,1,-1,0,0,0,0,0,1]$ corresponding to the cycles $\{e_2,e_3,e_4,e_9\}$
and $\{e_6,e_7,e_8\}$ and $\{e_1,e_2,e_3,e_4,e_{10}\}$ respectively.
\par 

%
%
\subsection{The Asymptotic Variance}\label{subsection:asymptotic}
Consider a time-invariant linear stochastic differential equation
with representation,
\begin{eqnarray*}
    \text{d} \mathbf{x}(t)
    & = & \mathbf{A} \mathbf{x}(t)  \text{d}t 
          + \mathbf{M} d \mathbf{v}(t), ~ \mathbf{x}(0) = \mathbf{x}_0, \\
        \mathbf{y}(t)
    & = & \mathbf{N} \mathbf{x}(t), 
\end{eqnarray*}
where $\mathbf{x}: \Omega \times T \rightarrow \mathbb{R}^{n_x}$; $\mathbf{A} \in \mathbb{R}^{n_x \times n_x}$;
 $\mathbf{M} \in \mathbb{R}^{n_x \times n_v}$; $\mathbf{v}: \Omega \times T \rightarrow \mathbb{R}^{n_v}$, is a standard Brownian motion with $\mathbf{v}(t) - \mathbf{v}(s) \in G(0, \mathbf{I}_{n_v} (t-s)), \forall ~ t, ~ s \in T, ~ s< t$; $\mathbf{x}_0 \in G(0, \mathbf{Q}_{\mathbf{x}_0})$ with $\mathbf{Q}_{\mathbf{x}_0}\in \mathbb{R}_{spd}^{n_x \times n_x}$; $\mathbf{y}: \Omega \times T \rightarrow \mathbb{R}^{n_y}$, $\mathbf{N} \in \mathbb{R}^{n_y \times n_x}$. 
  A standard Brownian motion is a stochastic process 
which starts at $t=0$ with $\mathbf{v}(0) = \mathbf{0}$,
has independent increments,
and the probability distribution of each increment is specified by
$(\mathbf{v}(t) - \mathbf{v}(s)) \in G(0, ~ (t-s)\mathbf{I}_{n_v})$ for any $s, ~ t \in T$ with $s < t$, meaning that 
$(\mathbf{v}(t)-\mathbf{v}(s))$ has a Gaussian probability distribution with mean zero and variance $(t-s)\mathbf{I}_{n_v}$. 

It follows from 
\cite[Theorem 1.52]{linearOptimalSystems} and
\cite[Theorem 6.17]{karatzas:shreve:1988}
that the state process $\mathbf{x}$ and the output process $\mathbf{y}$
are Gaussian processes.
Denote then for all $t \in T$,
$\mathbf{x}(t) \in G(\mathbf{m}_x(t), ~ \mathbf{Q}_{x, tv}(t))$ with $\mathbf{Q}_{x, tv}(t)\in \mathbb{R}_{spd}^{n_x \times n_x}$ and 
$\mathbf{y}(t) \in G(\mathbf{m}_y(t), ~ \mathbf{Q}_{y, tv}(t))$ with $\mathbf{Q}_{y, tv}(t)\in \mathbb{R}_{spd}^{n_y \times n_y}$.
If in addition the matrix $\mathbf{A}$ is Hurwitz
then there exists an invariant probability distribution
of this linear stochastic system 
with the representation and properties
\begin{eqnarray*}
\mathbf{0}
    & = & \lim_{t \rightarrow \infty} ~ \mathbf{m}_x(t), ~
          \mathbf{0} = \lim_{t \rightarrow \infty} ~ \mathbf{m}_y(t), 
          \nonumber \\
        \mathbf{Q}_x 
    & = & \lim_{t \rightarrow \infty} ~ \mathbf{Q}_{x, tv}(t), 
          \mathbf{Q}_y = \lim_{t \rightarrow \infty} ~ \mathbf{Q}_{y, tv}(t), 
\end{eqnarray*}
where the variance matrix
\begin{eqnarray*}
\mathbf{Q}_x
    & = & \int_0^{+\infty} \exp(\mathbf{A} t) 
          \mathbf{M} \mathbf{M}^{\top}
          \exp(\mathbf{A}^{\top} t) 
          \text{d}t,~~\mathbf{Q}_y=\mathbf{N}\mathbf{Q}_x\mathbf{N}^{\top}.
\end{eqnarray*}
Here $\mathbf{Q}_x$ is the unique solution of the matrix equation
\begin{eqnarray}
 \mathbf{0}
    & = & \mathbf{A}  \mathbf{Q}_x 
          + \mathbf{Q}_x \mathbf{A}^{\top} 
          + \mathbf{M} \mathbf{M}^{\top}. 
          \label{eq:lyapunoveq}
\end{eqnarray}
One calls the matrix
$\mathbf{Q}_x$ the {\em asymptotic variance of the state process} and 
$\mathbf{Q}_y$ the {\em asymptotic variance of the output process} and the matrix equation (\ref{eq:lyapunoveq})
the {\em (continuous-time) Lyapunov equation}
for the asymptotic variance $\mathbf{Q}_x$.
Because the matrix $\mathbf{A}$ is assumed to be Hurwitz,
this equation has a unique solution
which can be computed by a standard iterative procedure.
In general the solution $\mathbf{Q}_x$ 
is symmetric and positive semi-definite.
If the matrix tuple $(\mathbf{A}, ~ \mathbf{M})$ is a controllable pair
then the matrix $\mathbf{Q}_x$ is positive definite,
denoted by $0 \prec \mathbf{Q}_x$.
These results may be found in
\cite[Theorem 1.53, Lemma 1.5]{linearOptimalSystems} and
\cite{karatzas:shreve:1988}.

\section{Problem Formulation and the Main Results}\label{sec:asymptoticvariance}
In this section, we formulate the problem and present the main results of this paper. 
The reader may find the proofs of the results
in Section~\ref{sec:proofs}.
\par 
The dynamics of a complex network of coupled phase oscillators are described
in the following definition. 
\begin{definition}\label{def:kuramotosystem} 
\par
Consider an undirected graph 
$\mathcal{G} = (\mathcal{V},\mathcal{E})$ 
with a set of $n \in \mathbb{Z}_+$ nodes denoted by $\mathcal{V}$ and 
a set of $m \in \mathbb{Z}_+$ edges or lines denoted by $\mathcal{E}$.
The system of coupled phase oscillators 
is described by the dynamics \cite{Skardal2014,RobustnessSynchrony,PhysRevLett.114.038701},
\begin{eqnarray}
        d_i \frac{\emph{d} \delta_i(t)}{\emph{d}t}
    & = &  \omega_i+ 
          \sum_{j=1}^{n} ~ K_{ij} \sin(\delta_j(t) - \delta_i(t)), ~
          \forall i \in \mathcal{V} = \mathbb{Z}_n,
          \label{eq:originalmodel} 
\end{eqnarray}
with
\begin{eqnarray}
        K_{ij}
    & = & \left\{
          \begin{array}{ll}
            > 0, & \mbox{if} ~\exists ~ (i, ~ j) \in \mathcal{E}, \\
            = 0, & \mbox{else,} 
          \end{array}
          \right. ~ \forall ~ i, ~ j \in \mathbb{Z}_n.  \nonumber
\end{eqnarray}
where 
$d_i \in \mathbb{R}_{+}$ denotes the response time of oscillator $i$; 
$\delta_i: T \rightarrow \mathbb{R}$ denotes the phase; 
$\omega_i \in \mathbb{R}$ denotes the natural frequency 
which is a parameter  of the system; 
$K_{ij}$ denotes
the coupling strength of the nodes $i$ and $j$.
\end{definition}
\par 
When the network is complete, 
that is, every two nodes are connected, and $d_i=1$ for all the nodes 
and $K_{ij}=K/n$ for all $(i,j)\in\mathcal{E}$ 
with  $K\in\mathbb{R}_{+}$,
the system becomes the \emph{Kuramoto Model}
\cite{Kuramotomodel}. 
In the differential equation of a power system
there are two types of state variables,
phase angles and frequencies.
System (\ref{eq:originalmodel}) is an abstraction of a power system
in which there is only one set of state variables.
The state variables are denoted by $\delta_i$
but that need not correspond to the phase angles of a power system.
\begin{definition}\label{Def_synchronous}
Define a {\em synchronous state}
of the networked system (\ref{eq:originalmodel}) 
as the vector $\mathbf{\delta}^*(t)=\widetilde{\mathbf{\delta}}+(\widetilde{\omega} t)\mathbf{1}_n \in \mathbb{R}^n$, 
which is a solution of the equation
\begin{eqnarray}\label{flows}
        d_i\widetilde{\omega}
    & = &  \omega_i +
          \sum_{j=1}^{n}~ K_{ij} ~ \sin(\widetilde{\delta}_j - \widetilde{\delta}_i ),\text{for}~i=1,\cdots,n
\end{eqnarray}
and $\widetilde{\delta}=\text{col}(\widetilde{\delta}_i)\in\mathbb{R}^n$  that satisfies $\widetilde{\delta}_i-\widetilde{\delta}_j=(\delta_i^*(t)-\delta_j^*(t))(\emph{mod}(2\pi))$ for all $(i,j)\in\mathcal{E}$. 
\end{definition}
\par
 By summing all the equations in (\ref{flows}), it yields that at the synchronous state
 \begin{align}\label{synchronizedfrequency}
 \widetilde{\omega}=\frac{\sum_i^n{\omega_i}}{\sum_i^n{d_i}}\in\mathbb{R}.
 \end{align}
 \par 
 The existence of a  synchronous state can typically be obtained 
by increasing the coupling strength $K_{ij}$ for all the lines
 to sufficiently high values \cite{DorflerCriticalcoupling}. 
Consider the following 
 domain 
 \begin{align}
 \mathbf{\Theta}=\big\{\mathbf{\delta}\in\mathbb{R}^n\big||\delta_i-\delta_j|<\pi/2,~(i,j)\in\mathcal{E}\big\}\label{securitydomain}
 \end{align}
where $\mathbf{\delta}=\text{col}(\delta_i)$. 
It has been shown that the synchronous state in this domain is unique and exponentially stable \cite{skar_uniqueness_equilibrium,Kuramotostability}. In addition, if the initial state lies in this domain, the state of the system will converge to the synchronous state
 in this domain \cite{skar_uniqueness_equilibrium,Kuramotostability}. 
Instead of the phase difference in (\ref{securitydomain}), the \emph{counterclockwise difference} between the phases on the unit circle
is defined for identifying the subsets of the $n$-torus where there exist at most one synchronous state,
 see \cite{Jafarpour} for details.   
 \par 
Due to continuous disturbances, the state 
 will fluctuate around 
the synchronous state. If the fluctuations is too large such the state exits from the domain $\mathbf{\Theta}$,
the synchronization may be lost. Below attention is restricted to the stable synchronous states in the domain $\mathbf{\Theta}$. 
The fluctuations of the phase differences 
$(\delta_i-\delta_j)$ with $(i,j)\in\mathcal{E}$ in (\ref{eq:originalmodel}) 
around a synchronuous state are the object of study in this paper.
If the fluctuations are small
then the system 
operates in a neighborhood of a synchronous state.
The derivation of the linearized system of (\ref{eq:originalmodel})
is briefly summarized below with an assumption for the synchronous state.
\begin{assumption}\label{assumption:networkedsystem}
Consider the system (\ref{eq:originalmodel}), assume that 
(1) the graph $\mathcal{G}$ is connected, hence $m \geq n-1$ holds
;
(2) there exists a stable synchronous 
state such that the phase differences $|\widetilde{\delta}_i-\widetilde{\delta}_j|<\pi/2$ for all $(i,j)\in\mathcal{E}$.
\end{assumption}
\par 
We denote $\Delta \delta_i(t)=\delta_i(t) - \delta_i^*,\forall ~ i \in \mathbb{Z}_n$. The {\em linearized system},
linearized around the considered synchronous state, 
is then derived 
\begin{eqnarray}\label{linearizedsystem}
        d_i \frac{\text{d}}{\text{d}t} \Delta \delta_i(t)
    & = & \sum_{j=1}^n ~ w_{ij}  ~
          (\Delta \mathbf{\delta}_j(t) - \Delta \mathbf{\delta}_i(t) ), ~
          \forall ~ i \in \mathbb{Z}_n.
\end{eqnarray}
where 
\begin{eqnarray}\label{def_aij}
w_{ij}
    & = & \left\{
          \begin{array}{ll}
            K_{ij} ~ \cos(\widetilde{\delta}_i-\widetilde{\delta}_j), 
              & \mbox{if} ~\exists ~ (i, ~ j) \in \mathcal{E}, \\
            0, 
              & \mbox{else.} 
          \end{array}
          \right.
\end{eqnarray}

%
\begin{remark}
Consider the following more general dynamics \cite{Skardal2014}, 
\begin{align*}
\frac{\text{d}\delta_i}{\text{d}t}=\omega_i-K\sum_{j=1}^nA_{ij}H(\delta_i-\delta_j), ~\text{for}~ i=1,\cdots,n, 
\end{align*}
where $K$ is the coupling strength of lines, $A_{ij}=1$ if nodes $i$ and $j$ is connected,  $A_{ij}=0$ otherwise, 
$H$ is a $2\pi$ periodic coupling function.  If the coupling strength is sufficient large, there exists a synchronous 
state for this system. Clearly, after the linearization of this system around 
the synchronous state,  the system (\ref{linearizedsystem}) can be obtained. 
Another case is the consensus protocol 
for a system of $n$ autonomous multi-agent, 
\begin{eqnarray*}
        \frac{\text{d}x_i(t)}{\text{d}t}
    & = & \sum_{j=1}^{n} ~ a_{ij}(x_j(t)-x_i(t)), ~ i\in\mathbb{Z}_n.
\end{eqnarray*}
where $x_i(t)\in\mathbb{R}$ is a state variable, 
$a_{ij}$ is the coupling strength of the agents. 
The basic task is to achieve a consensus on a common state, 
that is, all $x_i(t)$ should converge to a common value $\bar{x}$ 
as $t\rightarrow \infty$.  
It has been shown that for a connected graph $\mathcal{G}$ for the agents and for $a_{ij}>0$,
consensus can be established \cite{Dorfler2012}.  
Clearly this consensus protocol with constant coupling strength is the same as the linearized system (\ref{linearizedsystem}). 
The theoretical result of this paper presented below 
may be directly applied to these types of systems and 
the performance of the synchronization of the systems can be investigated similarly.  
\end{remark}
 \par
The linearized system (\ref{linearizedsystem}) is then made stochastic
by defining a linear stochastic differential equation
driven by a Brownian motion process according to
\begin{eqnarray}
        ~~~~d_i ~ \text{d}\Delta \delta_i(t)
     =  \sum_{j=1}^n ~ w_{ij} ~ (\Delta \delta_j(t) - \Delta \delta_i(t) )\text{d}t
          + b_i ~ \text{d}\text{v}_i(t), ~ \Delta \delta_i(0) = 0, ~
          \forall ~ i \in \mathbb{Z}_n.
          \label{eq:kuramotosystemlinearizedstoc}
\end{eqnarray}
In this equation,
$\text{v}_i: \Omega \times T \rightarrow \mathbb{R}$ 
denotes a standard Brownian motion process 
which is a model for the disturbance and
$b_i \in \mathbb{R}_{+}$ models the strength of the disturbance,
it can be compared to the standard deviation.
Note that the noise $\text{v}_i$ affects only node $i$.
The noise components $\text{v}_1, ~ \text{v}_2, ~ \ldots, ~ \text{v}_n$
are assumed to be independent.
In equation (\ref{eq:kuramotosystemlinearizedstoc})
the state variable $\Delta \delta_i(t)$ denotes 
the deviation of the phase $\delta_i(t)$
from the phase $\delta_i^*(t)$ in the synchronous state. 
\par
To investigate 
the fluctuations of the phase differences 
of the line $k$ which connects the node pair $(i, ~ j)$
around the synchronous state $\delta^*(t)$, 
define the $k$-th output of the system as
\begin{eqnarray}
        y_k(t)
    & = & \Delta \delta_j (t) - \Delta \delta_i (t),
          \forall k \in \mathbb{Z}_m,
          \label{eq:outputk}
\end{eqnarray}
where $k$ denotes the index of the line $e_k$ 
which connects the nodes $i$ and $j$. 
The variance of the output component $y_k$ 
characterizes the phase difference in line $e_k$ 
in case of stochastic disturbances. 
Here, the direction of
the line $e_k$ is from node $j$ to $i$. Note that, the direction of the lines may be specified arbitrarily because it has no impact on the following analysis of the variance of the phase difference. After the specification of the directions
of the lines, the incidence matrix $\mathbf{C}$ of the graph is determined in (\ref{IncidenceDef}).   
\par 
The stochastic linear system (\ref{eq:kuramotosystemlinearizedstoc}) is written 
in a compact form of vectors, 
\begin{subequations}\label{stochasticprocesses}
\begin{eqnarray}
        \text{d} \mathbf{\delta} (t)
    & = & - \mathbf{D}^{-1} \mathbf{L} ~ \mathbf{\delta}(t) ~ \text{d}t
          + \mathbf{D}^{-1} \mathbf{B} ~ \text{d} \mathbf{v}(t), ~
          \mathbf{\delta}(0) = \mathbf{\delta}_0 \in \mathbb{R}^n, \\
        \mathbf{y}(t)
    & = & \mathbf{C}^{\top} 
          \mathbf{\delta} (t),
          \end{eqnarray}
          \end{subequations}
where $\delta(t)=\text{col}(\Delta\delta_i(t))\in\mathbb{R}^n$, $\mathbf{D}=\text{diag} (d_i) \in \mathbb{R}^{n\times n}$, $\mathbf{L}= (l_{ij}) \in \mathbb{R}^{n\times n}$ is the Laplacian matrix of graph $\mathcal{G}$ with the
weight $w_{ij}$ for the line $(i,j)\in\mathcal{E}$ defined in (\ref{def_aij}).
$\mathbf{B}= \text{diag} (d_i) \in \mathbb{R}^{n\times n}$,~$\mathbf{v}(t)=\text{col}(\text{v}_i(t)) \in \mathbb{R}^n$,~$\mathbf{y}=\text{col}(y_k) \in \mathbb{R}^m$, $ \mathbf{C} =(C_{ik}) \in \mathbb{R}^{n\times m}$ is the incidence matrix 
as defined in (\ref{IncidenceDef}). As in subsection \ref{subsection:graph}, 
we also define a diagonal matrix $\mathbf{R}=\text{diag} (R_k)\in\mathbb{R}^{m\times m}$ with $R_{k} = w_{ij}$
 being the weight of line $e_k$ with $k=(i,j)\in\mathcal{E}$.     
 The properties of the matrices $\mathbf{L},\mathbf{C}$ and $\mathbf{R}$ can be found in Lemma \ref{lemma:laplacianmatrixproperties}.   
\par
It is well known that for a time-invariant linear stochastic system
there exists a unique solution which satisfies
the stochastic differential equation.
Though the analysis of the stochastic linear system is valid only for 
comparatively small disturbances, it still provides intuitive insights on the robustness of the coupled phase oscillators.
The problem of the characterization of the asymptotic variance of the stochastic linear system is described below.
\begin{problem}\label{problem}
Consider the stochastic linear system (\ref{stochasticprocesses}).
Determine an analytic expression of the asymptotic variance
of the output process $\mathbf{y}$ and
display how this variance depends on the parameters of the system.
\end{problem}
\par 
The theorem for the solution of Problem \ref{problem}
makes use of the properties and the notations in the following lemma.

\begin{lemma}\label{lemma:laplacianmatrixtransformed}
Consider the stochastic linear system (\ref{stochasticprocesses})
with Assumption \ref{assumption:networkedsystem} and matrices $\mathbf{L}$ and $\mathbf{D}$. If the weights $w_{ij}$ of all the lines
are strictly positive and the matrix $\mathbf{D}$ is strictly positive definite, then 
there exist matrices with the following decomposition,
\begin{eqnarray}
 \mathbf{U} \mathbf{\Lambda}_n \mathbf{U}^{\top}
    & = & \mathbf{D}^{-1/2}
            \mathbf{L}
          \mathbf{D}^{-1/2},\label{diagonal} 
\end{eqnarray}
where $\mathbf{\Lambda}_n=\text{diag}(\lambda_i) \in \mathbb{R}_{diag}^{n \times n}$ with $0=\lambda_1< \lambda_2\leq\cdots\leq \lambda_n$ and $\lambda_i\in\mathbb{R}$ for $i=1,\cdots,n$ being the eigenvalues of the matrix $\mathbf{D}^{-1/2}\mathbf{L}\mathbf{D}^{-1/2}$, $\mathbf{U}= \begin{bmatrix}
    \mathbf{u}_1&\mathbf{u}_2 & \mathbf{u}_3 & \ldots & \mathbf{u}_{n}
    \end{bmatrix} \in \mathbb{R}_{ortg}^{n \times n}$
    with $\mathbf{u}_i\in\mathbb{R}^n$ being the eigenvector corresponding to 
    the eigenvalue $\lambda_i$ for $i=1,\cdots,n$. 
In addition, $\mathbf{u}_1 =\tau \mathbf{D}^{1/2}\mathbf{1}_n$ where  $\tau\in\mathbb{R}$.
\end{lemma}
\par 
The matrix decomposition in this Lemma has also been presented in \cite{GlobalMetrics}.
For the asymptotic variance matrix of the stochastic system (\ref{stochasticprocesses}), we have the following theorem. 
\begin{theorem}\label{theoremmain0}
Consider the stochastic system (\ref{stochasticprocesses}) with Assumption \ref{assumption:networkedsystem} and
the notations in Lemma  \ref{lemma:laplacianmatrixtransformed}. 
The  asymptotic variance of the output process $\mathbf{y}$ can be 
computed by 
\begin{eqnarray}
  \mathbf{Q}_y
    & = & \mathbf{C}^{\top}\mathbf{D}^{-1/2}
              \mathbf{U}_2
                \mathbf{Q}_{n-1}
              \mathbf{U}_2^{\top}\mathbf{D}^{-1/2}
          \mathbf{C}.  \label{Qdelta} 
\end{eqnarray}
where $\mathbf{U}_2 
    = \begin{bmatrix}
      \mathbf{u}_2 & \mathbf{u}_3 & \ldots & \mathbf{u}_{n}
    \end{bmatrix}\in \mathbb{R}^{n \times (n-1)}$ and $\mathbf{Q}_{n-1} = (q_{ij}) \in \mathbb{R}_{spd}^{(n-1)\times (n-1)}$ is the unique solution of the Lyapunov equation,
\begin{eqnarray}\label{LyapunovQnplus1}
\mathbf{0}
    & = & - \mathbf{\Lambda}_{n-1} \mathbf{Q}_{n-1} 
          - \mathbf{Q}_{n-1} \mathbf{\Lambda}_{n-1} 
          + \mathbf{U}_2^\top\mathbf{D}^{-1/2}\mathbf{B}\mathbf{B}^\top\mathbf{D}^{-1/2}\mathbf{U}_2, ~
\end{eqnarray}
with $\mathbf{\Lambda}_{n-1}
          = \text{diag}( \lambda_2, ~ \lambda_3, ~ \ldots, ~ \lambda_n )\in\mathbb{R}_{diag}^{(n-1) \times (n-1)}$.
In addition, the matrix $\mathbf{Q}_{n-1}$ is solved from the Lyapunov equation as
\begin{eqnarray}
  q_{ij}
    & = & (\lambda_{i}+\lambda_{j})^{-1}
          \mathbf{u}_{i}^{\top}\mathbf{D}^{-1/2}
          \mathbf{B} \mathbf{B}^{\top}\mathbf{D}^{-1/2}
          \mathbf{u}_{j},  \forall ~ i, ~ j = 2, ~ \cdots, ~ n, ~
          i \neq j, \label{Q1}
\end{eqnarray}
and in particular, 
\begin{eqnarray}
        q_{ii}
    & = & \frac{1}{2} \lambda_{i}^{-1}
          \mathbf{u}_{i}^{\top}
            \mathbf{D}^{-1/2}
              \mathbf{B} \mathbf{B}^{\top}
            \mathbf{D}^{-1/2}
          \mathbf{u}_{i},~
          \forall ~ i = 2, ~ \cdots, ~ n; \label{Q2}
\end{eqnarray}

\end{theorem}
\par 
The diagonal elements of $\mathbf{Q}_y$ 
are the variances of the phase differences of the lines.  
The line with the largest value is the most vulnerable. 
Hence, the most vulnerable lines can be identified directly 
by the values of the diagonal elements of that asymptotic variance matrix.
\par
It follows from Eqs. (\ref{Qdelta}-\ref{Q2}) 
that if the eigenvalues of the Laplacian 
matrix $\mathbf{D}^{-1/2}\mathbf{L}\mathbf{D}^{-1/2}$ increase, 
the variances of the phase differences decrease; 
consequently, the robustness increases.
This finding is consistent with 
the finding of a corresponding investigation by a perturbation method 
using a newly defined performance metric~\cite{RobustnessSynchrony}. 
In that investigation the robustness is related to the Kirchhoff indices.
The trace of $\mathbf{Q}_y$ 
is the $\mathcal{H}_2$ norm of the system 
(\ref{stochasticprocesses})
which is often used to study the performance 
of the synchronization of complex networks \cite{optimal_inertia_placement,RobustnessSynchrony}. 
\par 
For the network of complex phase oscillators with an assumption which follows, we further 
obtain an explicit formula of the asymptotic variance matrix. 
\begin{assumption}\label{assumption:varianceomponentsidentical}
Consider the stochastic process (\ref{stochasticprocesses}).
Assume that there exists a real number $\beta \in \mathbb{R}_{+}$ such 
that $ \forall ~ i \in \mathbb{Z}_n, ~ b_i^2/d_i = \beta$.
\end{assumption}
\par 
With this assumption, it holds
$\beta \mathbf{I}_n =  \mathbf{D}^{-1/2} \mathbf{B} \mathbf{B}^{\top} \mathbf{D}^{-1/2}$, which leads to 
\begin{eqnarray*}
        \mathbf{U}^{\top} 
          \mathbf{D}^{-1/2} 
            \mathbf{B}
            \mathbf{B}^{\top} 
          \mathbf{D}^{-1/2} 
        \mathbf{U}
    & = & \mathbf{U}^{\top} 
          \beta \mathbf{I}_n
          \mathbf{U}
          = 
          \beta \mathbf{I}_n.
\end{eqnarray*}
In addition, we obtain an explicit formula of the variance matrix in the next theorem. 

\par 
\begin{theorem}\label{theoremmain}
Consider the stochastic system (\ref{stochasticprocesses}) 
with the Assumptions \ref{assumption:networkedsystem} and \ref{assumption:varianceomponentsidentical}.  
The asymptotic variance of the output process $\mathbf{y}$ 
satisfies
\begin{eqnarray}
        \mathbf{Q}_y
    & = & \frac{\beta}{2}\mathbf{R}^{-1/2}
          \big(\mathbf{I}_m
            -\sum_{i=1}^{m-n+1} ~ \mathbf{X}_i\mathbf{X}_i^{\top}
          \big)
          \mathbf{R}^{-1/2}.
          \label{theoremmain-1}
\end{eqnarray}
where the vectors, $\mathbf{X}_i$ for $i=1,\cdots,m-n+1$, are a set of orthonormal basis vectors of the 
kernel of $\mathbf{C} \mathbf{R}^{1/2}$.
\end{theorem}
\par 
For graphs with cycles, the orthonormal basis vectors of the kernel of $\mathbf{C} \mathbf{R}^{1/2}$ is related 
to the cycle space of the graph $\mathcal{G}$ defined in Definition \ref{def:cyclespaceofgraph}. This relation and the method for computing the orthonormal basis vectors is 
explained in the following remark. 
\begin{remark}\label{orthonormalbasis}
Based on Theorem \ref{theoremcyclespace}, the vectors $\mathbf{\xi}_c$ for $c=1,\cdots,m-n+1$ obtained from the $m-n+1$ fundamental cycles
of the graph $\mathcal{G}$ form a set of basis vectors
of the kernel of the matrix $\mathbf{C}$. Thus, with the non-singular matrix $\mathbf{R}$, 
the set of vectors $\mathbf{R}^{-1/2} \mathbf{\xi}_c$ 
for $c=1,\cdots,m-n+1$, 
form a set of basis vectors of the kernel of 
$\mathbf{C} \mathbf{R}^{1/2}$. A set of orthonormal basis vectors 
of the kernel of $\mathbf{C} \mathbf{R}^{1/2}$ can then be derived 
by the Gram-Schmidt orthogonalization procedure applied to
the basis vectors $\mathbf{R}^{-1/2} \mathbf{\xi}_c$ for $c=1,\cdots,m-n+1$. 
Due to the non-uniqueness of the basis vectors $\mathbf{\xi}_c$ for $c=1,\cdots,m-n+1$ of the kernel of $\mathbf{C}$, the set of the orthonormal 
basis vectors of the kernel of $\mathbf{C} \mathbf{R}^{1/2}$ is also non-unique. However, for the kernel, a set of othonormal 
basis vectors can be obtained from another such set by a linear transformation consisting of an orthogonal matrix. Such a transformation does not influence the calculation of the variance in (\ref{theoremmain-1}).
\end{remark}

\par 
Formula (\ref{theoremmain-1}) 
shows explicitly the relation between the system parameters
and the asymptotic variance of the lines.
The variance increases linearly with respect to the factor $\beta$. 
Because vector $\mathbf{X}_i$ also depends 
on the weight $w_{ij}$ of the lines, 
the relationship between the variance and the weight is nonlinear. 
Furthermore, formula (\ref{theoremmain-1}) 
relates the robustness of the system 
to the cycle space of graphs. 
\emph{To the best knowledge of the authors, this is the first time that this relation is shown, which 
is important for studying the impact of the network topology on the synchronization stability.}
We remark that 
the cycles also play a role in the existence of the synchronous state \cite{multistability-cycles,Cycle-space1}. 
To illustrate the effects 
of increasing the coupling strength of a line or
of adding a new line using the theory of the cycle space,  
we introduce two concepts for graphs.
\begin{definition}\label{def:cyclecluster}
Consider a connected and undirected graph $\mathcal{G}$. 
\begin{enumerate}
\item[(i)] A single line is defined as an acyclic line that does not belong to any cycle;
\item[(ii)] Lines $e_1, e_2$ are cycle-shared if there exists at least one cycle containing both $e_1$ and $e_2$;
\item[(iii)] A cycle-cluster is a subgraph of $\mathcal{G}$ obtained in the following way. One starts from
a subgraph of one cycle and extends it by adding the lines in all the cycles with which the subgraph 
has at least one line in common, then one obtains a cycle-cluster.   
\end{enumerate}
%
%
\end{definition}
\par 
It is deduced from Definition \ref{def:cyclecluster} that a line either belongs to a cycle-cluster or is a single line and in a cycle-cluster, all the lines are in cycles and each pair of lines are cycle-shared lines. 
Taking the networks in Fig.\ref{fig1} as examples, it is seen that there 
are no cycle-clusters in network (a) and two cycle-clusters in networks (b) and (c).
All the lines in network (a) and lines $e_1$ and $e_5$
in network (b) and line $e_5$ in network (c) are singles lines. 
In network (c), each pair of lines in $\{e_1,e_2,e_3,e_4,e_9,e_{10}\}$ are cycle-shared lines
while $e_1,e_7$ are not cycle-shared because they are not contained in any cycles. See Table \ref{table0} for 
the details of the cycle-clusters and singles. 
\par
The next corollary of Theorem \ref{theoremmain} describes explicitly
the effects on the variances of the phase differences
of constructing a new line and
of increasing the coupling strength of a line.  The proof of this corollary 
makes use the theory of the cycle space, which can be found in Section \ref{sec:proofs}. 
\begin{table}[t]
\centering
\caption{The cycle-clusters and single lines of the networks in Fig \ref{fig1}. }
\begin{tabular}{|c |c| c|}
\hline
Network & line sets of cycle-clusters & single lines  \\
\hline
(a)& -- &$e_1-e_7$\\
(b)&$(e_2,e_3,e_4,e_9)$,~$(e_6,e_7,e_8)$&$e_1,e_5$\\
(c)&$(e_1,e_2,e_3,e_4,e_9,e_{10})$,~$(e_6,e_7,e_8)$&$e_5$\\
\hline
\end{tabular}
\label{table0}
\end{table}

\begin{corollary}\label{corollarymain}
Consider the stochastic system (\ref{stochasticprocesses}) with Assumption~\ref{assumption:varianceomponentsidentical}.
The following conclusions hold.
\begin{itemize}
\item [(i)]
The variance of the phase difference 
in a single-line connecting node $i$ and $j$ 
equals $\frac{\beta}{2}w_{ij}^{-1}$.
\item[(ii)]
Increasing the coupling strength of a single line does not affect the variances of the phase differences
in the other lines, and increasing the coupling strength of a line or constructing 
a new line in a cycle-cluster does not affect the variances of the phase differences of the lines that 
are not in this cycle-cluster. 
%
%
\item [(iii)]
If the coupling strength of lines are increased 
or new lines are constructed in a cycle-cluster, which does not change the weights of the other lines at the synchronous state,
then the variances of the phase differences in all the lines in this cycle cluster decrease.
%
\item [(iv)]
For a cycle-cluster 
with only one cycle with lines in the set $\mathcal{E}_c$ of the graph, 
the variance of the phase differences 
in the line connecting nodes $i$ and $j$ in this cycle-cluster equals
\begin{align}\label{eq:singlecycleone}
    \frac{\beta}{2}
    \Big(
      w_{ij}^{-1} 
      - w_{ij}^{-2}
      \Big(
        \sum_{(r,q)\in\mathcal{E}_c} ~
        w_{rq}^{-1}
      \Big)^{-1}
    \Big).
\end{align}
In addition, if it holds that $w_{ij}=\gamma$ for all the lines in this cycle, the variance becomes $\frac{\beta}{2\gamma}(1-\frac{1}{N})$ where $N$ is
the length of this cycle.
\end{itemize}
\end{corollary}
\par 
It is remarked that after increasing the coupling strength 
of a line or constructing a new line, if the phase differences 
in the other lines at the synchronous state are not changed, the weight of these lines will not change, in which 
case Corollary \ref{corollarymain}(iii) holds.
\par
Next the case is treated in which
Assumption~\ref{assumption:varianceomponentsidentical} does not hold, where bounds of the asymptotic variance are obtained.
\begin{theorem}\label{matrixbounds}
Consider the stochastic system (\ref{stochasticprocesses}).
Denote 
\begin{eqnarray*}
\underline{\beta}
     =  \min \{ b_i^2/d_i \in \mathbb{R}_{+} ~ 
                  \forall ~ i \in \mathbb{Z}_n 
               \}, 
        \overline{\beta} 
     =  \max \{ b_i^2/d_i \in \mathbb{R}_{+} ~ 
                  \forall ~ i \in \mathbb{Z}_n \}.
\end{eqnarray*}
The variance matrix $\mathbf{Q}_y$ 
of the phase differences in the lines satisfies
\begin{eqnarray}\label{bounds}
     \frac{1}{2} ~
          \underline{\beta} ~\widehat{\mathbf{Q}}
     \preceq \mathbf{Q}_y \preceq 
             \frac{1}{2} ~
             \overline{\beta} ~\widehat{\mathbf{Q}},~~
             \widehat{\mathbf{Q}}=
             \mathbf{R}^{-1/2} ~
             \big(
               \mathbf{I}_m
               - \sum_{i=1}^{m-n+1} ~
                   \mathbf{X}_i \mathbf{X}_i^{\top}
             \big) ~
             \mathbf{R}^{-1/2}. 
\end{eqnarray}
\end{theorem}
\section{Case study}\label{sec:examples}
In this section, by the three example networks shown in Fig. \ref{fig1}, we verify the explicit formula (\ref{Qdelta}) of the variance matrix and illustrate the findings in Corollary \ref{corollarymain} for the 
networks with Assumption \ref{assumption:varianceomponentsidentical}. We also verify the bounds of the variance matrix for the network without Assumption \ref{assumption:varianceomponentsidentical}. 
\begin{example}
Consider the three networks (a), (b), and (c)
with eight nodes displayed in Fig. \ref{fig1}. 
Networks (b) and (c) are constructed based on network (a) 
by adding lines $e_8,e_9$ and by adding lines $e_8,e_9, e_{10}$, respectively.  
We set $b_i=0.1$ 
for all the nodes in the 3 networks.  Three cases below are considered. 
\par 
\emph{Case 1}: $\omega_i=0,~d_i=1$ for all the nodes and $K_{ij}=1$ for all the lines in networks (a-c), thus $b_i^2/d_i=10^{-2}$ for all the nodes in the three networks;
\par 
\emph{Case 2}: $\omega_i=0,~d_i=1$ for all the nodes and $K_{ij}=1$ for the lines $e_1-e_9$ and $K_{{14}}=2$ for $e_{10}$ in network (c), thus $b_i^2/d_i=10^{-2}$ for all the nodes in this network. 
\par 
\emph{Case 3}: $\omega_i=0,~d_5=2, d_8=1/2$ and $d_i=1$ for all the other nodes
and $K_{ij}=1$ for the lines $e_1-e_9$ and $K_{{14}}=2$ for $e_{10}$ in network (c), thus 
$b_5^2/d_5=0.5\times 10^{-2}$, $b_8^2/d_8=2\times 10^{-2} $ and $b_i^2/d_i=10^{-2}$ for the other nodes in this network. 
\end{example}
\par 
Due to the setting of $\omega_i=0$ for all the nodes, it holds that the weight $w_{ij}=K_{ij}$ for all lines in the three cases. Thus, increasing the coupling strength of lines 
or constructing new lines has no impact on the weights. It is deduced that Assumption \ref{assumption:varianceomponentsidentical} holds in the networks in Cases 1-2 while does not 
hold in the one in Case 3.  
For the case with $\omega_i\neq 0$, the variances of the phase differences can be obtained from (\ref{theoremmain-1}) after calculating $w_{ij}$ from (\ref{def_aij}) where $\delta_i^*$ is solved 
from (\ref{flows}). Here, to verify the formula (\ref{theoremmain-1}), the findings in Corollary \ref{corollarymain} and the bounds of the matrices, it is sufficient to study 
the cases with $\omega_i=0$ only. In these cases, either constructing a new line or increasing 
the strength of a line will not change the weight of the other lines. 
\par 
The variances in the lines are shown in Table \ref{table3}. 
Here, the variances shown in all the rows except the ones with ($c^*$) and ($c^+$) are calculated 
by formula (\ref{theoremmain-1}) according to the procedure 
for computing the kernel of $\mathbf{C}\mathbf{R}^{1/2}$ 
in Remark \ref{orthonormalbasis} and 
are verified by Matlab using formula (\ref{Qdelta}).
For example, the variances in the lines in network (a) 
can be calculated directly from $\frac{\beta}{2}\mathbf{R}^{-1}$ 
because the cycle space of a tree network is empty. 
In network (b), the bases of the kernel of the cycle space are
$\mathbf{\xi}_1=[0,0,0,0,0,-1,1,-1,0]^{\top}$ and $\mathbf{\xi}_2=[0,-1,1,-1,0,0,0,0,1]^{\top}$, which are orthogonal. 
By scaling the vectors $\mathbf{R}^{-1/2}\mathbf{\xi}_i$
for $i=1,2$ to unit length with $\mathbf{R} =\mathbf{I}_m$, 
we derive
$\mathbf{X}_1=[0,0,0,0,0,-1/\sqrt{3},1/\sqrt{3},-1/\sqrt{3},0]^{\top}$ and
$\mathbf{X}_2=[0,-1/2,1/2,-1/2,0,0,0,0,1/2]^{\top}$.
We obtain the variances of the phase differences 
using formula (\ref{theoremmain-1}). 
In contrast, the numbers in the row with (c*) 
in the table are calculated 
from the simulations of system (\ref{stochasticprocesses}) 
for network (c) in Case 2. 
The simulation is conducted via the 
Euler-Maruyama method \cite{numericalmethod} 
with time $T=10000$ and time step $\text{d}t=10^{-3}$. 
The numbers in the row with ($c^+$) in the table are calculated by Matlab using formula (\ref{Qdelta}).
Table \ref{table3} shows that the statistical values of the variances 
in the last row are very close to the analytical 
values for network (c) in Case 2. 
This verifies the correctness of formula (\ref{theoremmain-1}).
\par

\begin{table}[t]
\centering
\caption{The diagonal elements of $\mathbf{Q}_{y}/b_i^2$ 
for the networks of Cases (1-2) in Fig \ref{fig1}; (c)-L and (c)-U denote the lower and upper bounds in case 3.}
\scalebox{0.85}{
\begin{tabular}{|c|c|c|c|c|c|c|c|c|c|c|c|}
\hline
Case    & Net.  &$e_1$&$e_2$&$e_3$&$e_4$&$e_5$& $e_6$&$e_7$&$e_8$&$e_9$&$e_{10}$\\ [4pt]\hline
    \multirow{4}*{1}&
    (a) &0.500 & 0.500 &0.500&0.500&0.500&0.500&0.500&$-$&$-$&$-$ \\[4pt] \cline{2-12}
    &(b) & 0.500& 0.375 &0.375&0.375&0.500&0.333&0.333&0.333&0.375&$-$\\[4pt] \cline{2-12}
    &(c) & 0.318 & 0.364 &0.364&0.364&0.500&0.333&0.333&0.333&0.273&0.318 \\[4pt] \cline{1-12}
 \multirow{2}*{2}&
(c)&0.278 & 0.361 &0.361&10.361&0.500&0.333&0.333&0.333&0.250&0.194\\[4pt] \cline{2-12}
&(c*)&$0.278$&$0.365$&$0.368$&$0.365$&$0.501$&$0.330$&$0.330$&$0.328$&$0.250$&$0.194$\\[4pt]
\cline{1-12}
 \multirow{3}*{3}&
(c)-L&0.139 & 0.181 &0.181&0.181&0.250&0.167&0.167&0.167&0.125&0.097\\[4pt] \cline{2-12}
&($c^+$)&$0.278$&$0.357$&$0.311$&$0.304$&$0.487$&$0.541$&$0.334$&$0.533$&$0.249$&$0.193$\\[4pt]\cline{2-12}
&(c)-U&0.556 & 0.722 &0.722&0.722&1.000&0.667&0.667&0.667&0.500&0.389\\[4pt] \cline{2-12}
\cline{1-12}
\end{tabular}
}
\label{table3}
\end{table}

The effects of adding new lines and increasing
coupling strength are described next with the networks in Fig. \ref{fig1}.
For the analytic derivation, see Corollary \ref{corollarymain}.
\par
First, 
the variance of the phase difference 
in a single line connecting nodes $i$ and $j$ 
is $\frac{\beta}{2}w_{ij}^{-1}$. 
When considering the single lines shown in Table \ref{table0}, 
we find that the variances in these lines are all $\beta/2$, 
as shown in Table \ref{table3}. 
In addition, the variance in line $e_5$ is not affected 
by adding lines $e_8,e_9$ in network (b) or 
by adding lines $e_8,e_9,e_{10}$ in network (c).  
Similarly, increasing the coupling strength of $e_9$ in network (c) 
in Case 2 also has no impact on this variance.
\par
Second, 
adding new lines or increasing the coupling strength of lines 
in a cycle-cluster do not affect the variance of the phase differences 
in the lines outside this cycle-cluster. 
In network (c) of Case 1, 
the variances in lines $e_6,e_7,e_8$ 
are the same as those in network (b) and 
are not affected by adding of $e_{10}$. 
Similarly, in network (c) of Case 2, 
these variances are not changed by increasing the coupling strength 
of line $e_{10}$ from $1$ to $2$ 
because line $e_{10}$ is not in the cycle-cluster of $(e_6,e_7,e_8)$.
\par
Third, 
by adding new lines or increasing the coupling strength of lines 
in a cycle-cluster, 
the variances of the phase differences 
in all the lines of this cycle-cluster will decrease.
The calculation for networks (b-c) in Case 1 verify this finding, 
where the variances in lines $e_2,e_3,e_4,e_9$ 
decrease from $3\beta/8$ in network (b) 
to $4\beta/11$,
$4\beta/11$,
$4\beta/11$, and 
$3\beta/11$ in network (c), 
respectively, after adding line $e_{10}$. 
In addition, 
the variances further decrease to $13\beta/36,13\beta/36,13\beta/36$ and 
$\beta/4$ when the coupling strength of line $e_{10}$ 
increases from $1$ to $2$ in network (c) in Case 2.
\par
Fourth, 
for a cycle-cluster with only one cycle with lines in set $\mathcal{E}_c$ 
in the graph, the variance of the phase difference 
in the line connecting nodes $i$ and $j$ is
\begin{align}\label{eq:singlecycletwo}
    \frac{\beta}{2}
    \Big(
      w_{ij}^{-1} - w_{ij}^{-2}
      \big(
        \sum_{(r,q)\in\mathcal{E}_c} ~ w_{rq}^{-1}
      \big)^{-1}
    \Big).
\end{align}
In addition, if $w_{ij}=\gamma$ holds for all the lines, 
the variance becomes $\frac{\beta}{2\gamma}\big(1-\frac{1}{N}\big)$,
where $N$ is the length of the cycle. 
In network (b) in Case 1, there are two cycles. 
By means of (\ref{eq:singlecycletwo}), 
it is obtained that the variances in lines $e_6,e_7,e_8$ 
are all $\beta/3$ and those in lines $e_2,e_3,e_4,e_9$ are all $3\beta/8$. 
This result demonstrates that forming small cycles 
can effectively suppress the variances of the phase differences and the benefit of forming 
a cycle of length $N$ is of the order of $O(N^{-1})$. This is consistent 
with the findings in \cite{PhysRevE.99.062213,Xi2016}.
We conclude that formula (\ref{eq:singlecycletwo}) 
provides a conservative estimate of the variances 
in lines in cycle-clusters. 
In other words, the variance in a line that is in multiple cycles 
can be approximated by formula (\ref{eq:singlecycletwo}) 
by considering the smallest cycle that includes this line. 
For example, the variance in line $e_1$ in network (c) in Case 1 
can be approximated as $\beta/3$, 
which is calculated in the cycle $(e_1,e_9,e_{10})$ 
by formula (\ref{eq:singlecycletwo}) and 
is slightly larger than $7\beta/22$, 
as shown in Table \ref{table3}. 
Clearly, this value is conservative.
\par
Finally, 
increasing the scale of the network by adding nodes with $b_i^2/d_i=\beta$ will neither decrease nor increase 
the fluctuations of the phase differences; 
thus, it has no impact on the synchronization stability. 
This is a result from formula (\ref{theoremmain-1}), 
where the variance matrix is determined 
by the strength of disturbance, the cycle space and the weights of the lines.
\par
Regarding the vulnerability, 
we find that based on formula (\ref{theoremmain-1}) single lines are usually the most vulnerable lines 
which are the bottleneck 
of the network on the synchronization stability.
We remark that for the networks with non-uniform ratio $b_i^2/d_i$ among the nodes,  
the lines with the most serious fluctuations 
can be identified by formula (\ref{Qdelta}).
With respect to the bounds of the variances for the 
networks with non-uniform ratio $b_i^2/d_i$, it is seen that all the values of the variance 
are constrained by the  lower and upper bounds.  In addition, when comparing the 
variance in Case 3 with those in Case 2, it is found that the variances in 
the lines $e_1-e_5, e_9-e_{10}$ all decrease mainly due to the increase of the damping coefficient 
at node $5$ while those in the lines $e_6-e_8$ increase mainly due to the 
decrease of the damping coefficient at node $8$.  This indicates that the variances 
of the lines in a cycle-cluster are mainly influenced by the disturbances at the nodes in this 
cycle cluster.  We remark that the increase of $d_5$ also influences the variances in 
lines $e_6-e_8$ which however is overtaken by the decrease of $d_8$ in the cycle-
cluster $\{e_6,e_7,e_8\}$.
\section{The Proofs}\label{sec:proofs}

\begin{proof}

Proof of Lemma 3.6.
\par 
The spectral decomposition of matrix $\mathbf{D}^{-1/2}\mathbf{L}\mathbf{D}^{-1/2}$
follows directly from the positive definiteness of the diagonal matrix $\mathbf{D}$ and symmetric-semi-positive-definiteness of the Laplacian matrix $\mathbf{L}$ which has a zero eigenvalue and $n-1$ positive real eigenvalues.
Because $\tau\mathbf{1}_n$ is 
the eigenvector corresponding to the zero eigenvalue of the matrix $\mathbf{L}$, $\tau\mathbf{D}^{1/2}\mathbf{1}_n$ 
is the eigenvector related to the zero eigenvalue of the matrix $\mathbf{D}^{-1/2}\mathbf{L}\mathbf{D}^{-1/2}$.
\end{proof}

\begin{proof}
Proof of Theorem~\ref{theoremmain0}.
\par
Consider the stochastic system (\ref{stochasticprocesses}). 
Let $ \mathbf{x}(t) =  \mathbf{U}^{\top} \mathbf{D}^{1/2} \mathbf{\delta}(t), \mathbf{x}: \Omega \times T \rightarrow \mathbb{R}^n$, transform the stochastic differential equation according 
to Lemma~ \ref{lemma:laplacianmatrixtransformed}, 
\begin{eqnarray*}
 \text{d}\mathbf{x}(t)
    & = & - \mathbf{\Lambda}_n \mathbf{x}(t) \text{d}t
          + \mathbf{U}^{\top} \mathbf{D}^{-1/2} \mathbf{B} \text{d}\mathbf{v}(t).
\end{eqnarray*}
where the formula  below is applied,
\begin{eqnarray*}
 \mathbf{U}^{\top} \mathbf{D}^{1/2} \mathbf{D}^{-1} 
            \mathbf{L} \mathbf{D}^{-1/2} \mathbf{U}
          = \mathbf{U}^{\top} \mathbf{D}^{-1/2} 
              \mathbf{L} \mathbf{D}^{-1/2} \mathbf{U} 
          = \mathbf{\Lambda}_n. 
\end{eqnarray*}

Decompose this stochastic differential equation
according to the formulas
\begin{eqnarray*}
    \mathbf{x}(t)
    & = &\begin{bmatrix}
     x_1(t) \\ \mathbf{x}_2(t)
    \end{bmatrix},~
          x_1: \Omega \times T \rightarrow \mathbb{R}^1, ~
          \mathbf{x}_2: \Omega \times T \rightarrow \mathbb{R}^{n-1}, 
\end{eqnarray*}
and 
\begin{eqnarray}
        \mathbf{\Lambda}_n
    & = &
    \begin{bmatrix}
     0                & \mathbf{0}_{n-1}^{\top} \\
            \mathbf{0}_{n-1} & \mathbf{\Lambda}_{n-1}
    \end{bmatrix}\in \mathbb{R}^{n \times n}, ~
          \mathbf{\Lambda}_{n-1} \in \mathbb{R}^{(n-1) \times (n-1)}, 
          \label{eq:lambdan} 
\end{eqnarray}
we obtain
\begin{eqnarray}
        \text{d}\mathbf{x}_2(t)
    & = & - \mathbf{\Lambda}_{n-1} \mathbf{x}_2(t) \text{d}t 
          + \mathbf{U}_2^{\top} \mathbf{D}^{-1/2} \mathbf{B} \text{d} \mathbf{v}(t), 
\end{eqnarray}
and the output becomes
\begin{eqnarray}
        \mathbf{y}(t)
    & = & \mathbf{C}^{\top} \mathbf{\delta}(t)
          = \mathbf{C}^{\top} 
              \mathbf{D}^{-1/2} 
            \mathbf{U} \mathbf{x}(t)
          =\begin{bmatrix}
           \mathbf{C}^\top\mathbf{D}^{-1/2} \mathbf{u}_1 & \mathbf{C}^{\top} \mathbf{D}^{-1/2}\mathbf{U}_2
          \end{bmatrix}
           \mathbf{x}(t), ~ \nonumber \\
    &   & \mbox{using Lemma~\ref{lemma:laplacianmatrixtransformed}
             and Lemma~\ref{lemma:laplacianmatrixproperties} one obtains
          } \nonumber \\
    & = & \begin{bmatrix}
       0 & \mathbf{C}^{\top}\mathbf{D}^{-1/2} \mathbf{U}_2
    \end{bmatrix}
    \begin{bmatrix}
              x_1(t) \\ \mathbf{x}_2(t)
\end{bmatrix}
            = \mathbf{C}^\top\mathbf{D}^{-1/2}  \mathbf{U}_2\mathbf{x}_2(t).
\end{eqnarray}
\par
From the summary of properties of linear stochastic differential equations in subsection \ref{subsection:asymptotic}, it 
follows that both $\mathbf{x}_2$ and $\mathbf{y}$
are stationary Gausian stochastic processes
of which the asymptotic variance matrices are determined by the equations
(\ref{LyapunovQnplus1}) and (\ref{Qdelta}) respectively. 
It follows from consideration of $- \mathbf{L}$,
from Lemma~\ref{lemma:laplacianmatrixtransformed}
that the matrix $-\mathbf{\Lambda}_{n-1}$ is a Hurwitz matrix.
Hence there exists a unique solution $\mathbf{Q}_{n-1}$ of the above
Lyapunov equation which is positive semi-definite.
The formula for the variance of the output process
$\mathbf{y}$ follows then from the equation of the output.
\par
The analytic form of the equations (\ref{Q1}-\ref{Q2})
then follows directly from the above Lyapunov equation
using the fact that the matrix $\mathbf{\Lambda}_{n-1}$
is a diagonal matrix.
\end{proof}
\par
Before introducing the proof of Theorem \ref{theoremmain}, 
another lemma is presented. 
\begin{lemma}
Define the matrix
$\widehat{\mathbf{C}} 
= \mathbf{R}^{1/2} \mathbf{C}^{\top} \mathbf{D}^{-1/2} \in \mathbb{R}^{m \times n}$.
If $m \geq n$, then there exists an orthogonal matrix 
$\mathbf{W} \in \mathbb{R}^{m\times m}$ such that 
\begin{eqnarray}
        \mathbf{W}^{\top} \widehat{\mathbf{C}} 
        \widehat{\mathbf{C}}^{\top} \mathbf{W}
    & = & \mathbf{\Lambda}_m, \label{lemma2-1} 
\end{eqnarray}
where
\begin{eqnarray*}
        \mathbf{\Lambda}_m
    & = & \begin{bmatrix}
            \mathbf{\Lambda}_{n-1} & \mathbf{0}\\
            \mathbf{0}             & \mathbf{0} 
          \end{bmatrix} \in \mathbb{R}^{m\times m}. ~
\end{eqnarray*}
Denote
\begin{eqnarray*}
        \mathbf{W} 
    & = & \begin{bmatrix}
            \mathbf{w}_1 & \mathbf{w}_2 & \cdots 
              & \mathbf{w}_{n-1} & \mathbf{w}_n & \cdots 
              & \mathbf{w}_m
            \end{bmatrix}.
\end{eqnarray*}
The vector $\mathbf{w}_i$ 
is an orthonormal eigenvector of 
$\widehat{\mathbf{C}}\widehat{\mathbf{C}}^{\top}$ 
corresponding to the nonzero eigenvalue 
$\lambda_{i+1}$ for $i=1,\cdots,n-1$ and 
$\mathbf{w}_i$ for $i=n,\cdots,m$ 
are the orthonormal eigenvectors corresponding to the zero eigenvalues.
\end{lemma}
\begin{proof}
For the connected graph $\mathcal{G}$, 
it holds that $\rank(\mathbf{C}) = n-1$, 
which leads to $\rank(\widehat{\mathbf{C}}) = n-1$. 
Because the kernel of $\widehat{\mathbf{C}} \widehat{\mathbf{C}}^{\top}$ and 
$\widehat{\mathbf{C}}^{\top}$ are identical, 
we obtain $\rank(\widehat{\mathbf{C}}\widehat{\mathbf{C}}^{\top}) = n-1$.
Hence, it only needs to be proven that
the non-zero diagonal elements of $\mathbf{\Lambda}_m$ 
are the non-zero eigenvalues of 
$\widehat{\mathbf{C}} \widehat{\mathbf{C}}^{\top}$.
We obtain from 
equations (\ref{incidenceLap}) and (\ref{diagonal}) 
that 
\begin{eqnarray*}
        \mathbf{U}^{\top} \widehat{\mathbf{C}}^{\top} \widehat{\mathbf{C}}
        \mathbf{U}
    & = & \mathbf{U}^{\top}
          \mathbf{D}^{-1/2}
          \mathbf{C}
          \mathbf{R}
          \mathbf{C}^{\top}
          \mathbf{D}^{-1/2}
          \mathbf{U}
          = \mathbf{U}^{\top}
          \mathbf{D}^{-1/2}
          \mathbf{L}
          \mathbf{D}^{-1/2}
          \mathbf{U}
          = \mathbf{\Lambda}_n.
\end{eqnarray*} 
We premultiply the above equation by 
$\widehat{\mathbf{C}} \mathbf{U}$, then derive 
\begin{equation}\label{eigenequation}
      \widehat{\mathbf{C}} 
      \widehat{\mathbf{C}}^{\top}
      \widehat{\mathbf{C}}
      \mathbf{U}
    = \widehat{\mathbf{C}}
      \mathbf{U}
      \mathbf{\Lambda}_n.
\end{equation}
As in Lemma~\ref{lemma:laplacianmatrixtransformed}, we write $\mathbf{U}$ into  
$\begin{bmatrix}
  \mathbf{u}_1 & \mathbf{u}_2 & \mathbf{u}_3 & \cdots & \mathbf{u}_n
\end{bmatrix}$ with 
$\mathbf{D}^{-1/2}\mathbf{u}_1 = \tau\mathbf{1}_n$. 
It follows from 
$\mathbf{C}^{\top} \mathbf{1}_n = \mathbf{0}$  that $\widehat{\mathbf{C}} \mathbf{u}_1 = \mathbf{0}$. 
Hence, we obtain from (\ref{eigenequation}) that
\begin{eqnarray*}
        \widehat{\mathbf{C}} \widehat{\mathbf{C}}^{\top}
        \begin{bmatrix}
            \widehat{\mathbf{C}} \mathbf{u}_2 
          & \widehat{\mathbf{C}} \mathbf{u}_3
          & \cdots
          & \widehat{\mathbf{C}} \mathbf{u}_n
        \end{bmatrix}
    & = & \begin{bmatrix}
              \lambda_2 \widehat{\mathbf{C}} \mathbf{u}_2
            & \lambda_3 \widehat{\mathbf{C}} \mathbf{u}_2
            & \cdots
            & \lambda_n \widehat{\mathbf{C}} \mathbf{u}_n
          \end{bmatrix}
\end{eqnarray*}
which demonstrates that $\lambda_i$ 
is an eigenvalue of $\widehat{\mathbf{C}} \widehat{\mathbf{C}}^{\top}$ 
with corresponding eigenvectors 
$\widehat{\mathbf{C}} \mathbf{u}_i$ for $i=2, ~ \cdots, ~ n$. 
\end{proof}
\begin{proof}
Proof of Theorem~\ref{theoremmain}.
\par
With the assumption of 
$\mathbf{D}^{-1/2} \mathbf{B} \mathbf{B}^{\top} \mathbf{D}^{-1/2}
  = \beta \mathbf{I}_n
$, 
we obtain from (\ref{Q1}-\ref{Q2}) that 
\begin{equation}\label{Qmatrix}
    \mathbf{Q}_{n-1}
    = \frac{\beta}{2} \mathbf{\Lambda}_{n-1}^{-1}.
\end{equation}
\par
By the definition of the Moore-Penrose inverse of a symmetric matrix, 
we obtain from (\ref{eq:lambdan}) that 
\begin{eqnarray}
        (\mathbf{D}^{-1/2} \mathbf{L} \mathbf{D}^{-1/2})^\dag
    & = & \mathbf{U} \Lambda^\dag_n \mathbf{U}^{\top}
          = \mathbf{U}_2 \mathbf{\Lambda}_{n-1}^{-1}\mathbf{U}_2^{\top},
          \label{inverse}
\end{eqnarray}
where $(\cdot)^\dag$ denotes the Moore-Penrose inverse of a matrix. 
By inserting (\ref{Qmatrix}) into (\ref{Qdelta}), 
we obtain from (\ref{inverse}) that 
\begin{eqnarray*}
        \mathbf{Q}_{\mathbf{y}}
    & = & \frac{\beta}{2} 
          \mathbf{C}^{\top} \mathbf{D}^{-1/2}
          (\mathbf{D}^{-1/2} \mathbf{L} \mathbf{D}^{-1/2})^\dag
           \mathbf{D}^{-1/2} \mathbf{C}  \\
    & = & \frac{\beta}{2}
          \mathbf{C}^{\top}
          \mathbf{D}^{-1/2}
          (\mathbf{D}^{-1/2}
           \mathbf{C} \mathbf{R}
           \mathbf{C}^{\top}
           \mathbf{D}^{-1/2}
          )^\dag\mathbf{D}^{-1/2}\mathbf{C} 
          = \frac{\beta}{2} \mathbf{R}^{-1/2}\widehat{\mathbf{C}}
          \big(
            \widehat{\mathbf{C}}^{\top} 
            \widehat{\mathbf{C}}
          \big)^\dag \widehat{\mathbf{C}}^{\top} \mathbf{R}^{-1/2}.
\end{eqnarray*}
By left multiplying (\ref{lemma2-1}) 
with $\widehat{\mathbf{C}}^{\top}\mathbf{W}$, 
we obtain
\begin{eqnarray*}
        \widehat{\mathbf{C}}^{\top}
        \widehat{\mathbf{C}}
        \widehat{\mathbf{C}}^{\top}
        \mathbf{W}
    & = & \widehat{\mathbf{C}}^{\top}
          \mathbf{W}
          \mathbf{\Lambda}_m,
\end{eqnarray*}
which indicates that the column vectors of 
$\widehat{\mathbf{C}}^{\top}\mathbf{W}$ 
are the eigenvectors of $\widehat{\mathbf{C}}^{\top}\widehat{\mathbf{C}}$. 
We focus on the first $n-1$ eigenvectors 
$\widehat{\mathbf{C}}\mathbf{w}_1, ~ \cdots, ~
  \widehat{\mathbf{C}} \mathbf{w}_{n-1}$ 
in matrix $\widehat{\mathbf{C}}^{\top}\mathbf{W}$, which are orthogonal.
From the normalization of $\widehat{\mathbf{C}}\mathbf{w}_i$ 
for $i=1, ~ \cdots, ~ n-1$, it yields
\begin{eqnarray*}
    &   & \lambda_{2}^{-1/2}\widehat{\mathbf{C}}\mathbf{w}_1, ~
          \lambda_{3}^{-1/2}\widehat{\mathbf{C}}\mathbf{w}_2, ~
          \cdots, ~
          \lambda_{n}^{-1/2}\widehat{\mathbf{C}}\mathbf{w}_{n-1}.
\end{eqnarray*}
With these unit vectors, 
we derive that the Moore-Penrose inverse of 
$\widehat{\mathbf{C}}^{\top}\widehat{\mathbf{C}}$ satisfies
\begin{eqnarray*}
        \Big( 
          \widehat{\mathbf{C}}^{\top}\widehat{\mathbf{C}}
        \Big)^{\dag}
    & = & \sum_{i=2}^{n} ~
            {\lambda_i^{-2}
             (\widehat{\mathbf{C}} \mathbf{w}_{i-1}) ~
             (\widehat{\mathbf{C}} \mathbf{w}_{i-1})^{\top}
            }. 
\end{eqnarray*}
From (\ref{lemma2-1}), we further get
\begin{eqnarray*}
        \widehat{\mathbf{C}}
        \Big(
          \widehat{\mathbf{C}}^{\top}
          \widehat{\mathbf{C}}
        \Big)^{\dag}
        \widehat{\mathbf{C}}^{\top} 
    & = & \sum_{i=2}^n ~ 
          {
            \lambda_{i}^{-2}
            \widehat{\mathbf{C}}
            \widehat{\mathbf{C}}^{\top}
            \mathbf{w}_{i-1}
            \mathbf{w}_{i-1}^{\top}
            \widehat{\mathbf{C}}
            \widehat{\mathbf{C}}^{\top}
          } 
          = \sum_{i=2}^{n} ~ 
          \mathbf{w}_{i-1}
          \mathbf{w}_{i-1}^{\top} 
     =  \mathbf{I}_m 
          - \sum_{i=n}^{m} ~ {\mathbf{w}_i \mathbf{w}_i^{\top}}.
\end{eqnarray*} 
Here $\mathbf{w}_i$ for $i=n, ~ \cdots, ~ m$ 
are the orthonormal eigenvectors 
corresponding to the zero eigenvalue such that 
$\mathbf{w}_i^{\top} \widehat{\mathbf{C}} 
  \widehat{\mathbf{C}}^{\top} \mathbf{w}_i=0$, 
thus $\widehat{\mathbf{C}}^{\top}\mathbf{w}_i = 0$. 
From $\widehat{\mathbf{C}}^{\top} 
  = \mathbf{D}^{-1/2} \mathbf{C} \mathbf{R}^{1/2}$, 
we obtain $\mathbf{C} \mathbf{R}^{1/2} \mathbf{w}_i = 0$,
which indicates that the vectors $\mathbf{w}_i$ for $i = n, ~ \cdots, ~ m$ 
form an orthonormal basis of the kernel of $\mathbf{C} \mathbf{R}^{1/2}$. 
We denote
$\mathbf{X}_i = \mathbf{w}_{i+n-1}$ for $i=1, ~ \cdots, ~ m-n+1$, 
which completes the proof. 
\end{proof}
\begin{proof}
Proof of Corollary~\ref{corollarymain}.
\par 
(i) If the network is acyclic, all the lines are single-lines. 
There are no non-zero 
elements in the cycle space, 
the variance matrix of the phase difference is thus 
$\frac{\beta}{2} \mathbf{R}^{-1}$ 
obtained from (\ref{theoremmain-1}).
If there are cycles in the network, 
without loss of generality, 
assume line $e_1$ is a single-line. 
By the method to formulate the basis of the cycle space in subsection \ref{subsection:graph}, 
the basis of the cycle space has the form 
$\mathbf{\xi}_i=
  \begin{bmatrix}
    0 & \xi_{i,2} & \xi_{i,3} & \cdots & \xi_{i,m}
  \end{bmatrix}^{\top}
$
where $\xi_{i,j}$ is either $-1$, $1$, or $0$, and $\mathbf{X}_i$ 
can be obtained by the Gram-Schmidt orthogonalization 
of $\mathbf{R}^{-1/2} \mathbf{\xi}_i$, 
which has the form 
$\mathbf{X}_i=
  \begin{bmatrix}
    0 & x_{i,2} & x_{i,3} & \cdots & x_{i,m}
  \end{bmatrix}^{\top}
$.
Because the elements in the first column and 
the first row of $\mathbf{X}_i \mathbf{X}_i^{\top}$ 
are all zero, 
we obtain from (\ref{theoremmain-1}) 
that the variance of the phase differences in this line 
is $ \frac{\beta}{2w_{ij}}$. 
\par 
(ii)Without loss of generality,  assume there are three sub-graphs in  $\mathcal{G}=(\mathcal{V},\mathcal{E})$, i.e., $\mathcal{G}_1=(\mathcal{V}_1,\mathcal{E}_1)$, $\mathcal{G}_2=(\mathcal{V}_2,\mathcal{E}_2)$ and $\mathcal{G}_3=(\mathcal{V}_3,\mathcal{E}_3)$,
where $\mathcal{G}_1$ is either  a cycle-cluster or a single line,  $\mathcal{E}_1\cup\mathcal{E}_2\cup\mathcal{E}_3=\mathcal{E}$,  $\mathcal{E}_i\cap\mathcal{E}_j=\emptyset$ for $i\neq j$, and $\mathcal{V}_1\cap \mathcal{V}_2=\{p\}$, $\mathcal{V}_1\cap \mathcal{V}_3=\{q\}$
and $\mathcal{V}_2\cap \mathcal{V}_3=\emptyset$. Here $p,q$ denote the index of two nodes respectively. We prove that
increasing the coupling strength of a line in $\mathcal{G}_1$ or constructing a new line in $\mathcal{G}_1$ has no impact on the variances of the phase differences in the lines
in $\mathcal{G}_2$ and $\mathcal{G}_3$. Here, we say constructing a new line in $\mathcal{G}_1$ only when $\mathcal{G}_1$ is a cycle-cluster. 
\par 
From the formula (\ref{theoremmain-1}), it is seen that the variance depends on the phase differences at the synchronous state, which play a role in the terms $\mathbf{R}$ and $\mathbf{X}_i$.  Due to the dependence of the phase differences at the synchronous state on the network topology, constructing new lines or increasing the coupling strength affect the variance in a non-linear way. Hence, we prove this conclusion in two steps.
\par 
First, we prove that the phase differences at the synchronous state in the lines of $\mathcal{G}_2$
are independent of constructing new lines and increasing the coupling strength of lines in  $\mathcal{G}_1$ and similarly for $\mathcal{G}_3$. From (\ref{synchronizedfrequency}), it is seen that the synchronized 
frequency $\widetilde{\omega}$ is not affected by these actions.  With (\ref{synchronizedfrequency}),
it is deduced that the sum of the equations in (\ref{flows}) is zero, which indicates the equations are singular and 
the phases at the nodes cannot be solved directly.  To obtain the phase differences at the synchronous state, a node has to be selected as the reference node at which the phase is zero. The selection of the reference 
node does not affect the phase difference due to the uniqueness of the synchronous state with Assumption \ref{assumption:networkedsystem}.  The common node $p$ of $\mathcal{G}_1$ and $\mathcal{G}_2$ is selected as 
the reference node. Then we obtain the equations for the calculation of the phase differences in $\mathcal{G}_2$,
\begin{align*}
        d_i\widetilde{\omega}
    & =   \omega_i +
          \sum_{j\in\mathcal{V}_2}~ K_{ij} ~ \sin(\widetilde{\delta}_j - \widetilde{\delta}_i ),\text{for}~i\in\mathcal{V}_2~\text{and}~i\neq p\\
          \widetilde{\delta}_p&=0. 
\end{align*}
Clearly, these equations are not affected by either increasing the coupling strength of lines or 
constructing new lines in $\mathcal{G}_1$.  Thus, the phase differences solved from the above equations 
are not affected by these changes.  It follows that the weights $w_{ij}$ of the lines in $\mathcal{G}_2$ are also not affected.  
\par 
Second, we prove that the variances of the phase differences in the lines of $\mathcal{G}_2\cup\mathcal{G}_3$
 are not influenced by constructing a new line or increasing the coupling strength of a line
 in $\mathcal{G}_1$.  (i) If graph $\mathcal{G}$ is a tree, then $\mathcal{G}_1$ is a single line and the conclusion is obtained from (a) directly due to the unchanged weights of the lines in $\mathcal{G}_2\cup\mathcal{G}_3$. 
  (ii) If there are cycles in $\mathcal{G}$ and $\mathcal{G}_1$ is a single line, 
  which is denoted by $e_1$, then
 the basis vectors for the fundamental cycles of $\mathcal{G}$ have the form 
$\mathbf{\xi}_i=
  \begin{bmatrix}
    0 & \xi_{i,2} & \xi_{i,3} & \cdots & \xi_{i,m}
  \end{bmatrix}^{\top}
$ for $i = 1, ~ \cdots, ~ m-n+1$, where 
$\xi_{i,q} = 1, ~ -1, ~\text{or,} ~ 0$ for $q=2, ~ \cdots, ~ m$.  
By the Gram-Schmidt orthogonalization of $\mathbf{R}^{-1/2}\xi_i$, 
we obtain
$\mathbf{X}_i=
  \begin{bmatrix}
    0 & x_{i,2} & x_{i,3} & \cdots & x_{i,m}
  \end{bmatrix}^{\top}
$ 
for $i = 1, ~ \cdots, ~ m-n+1$ 
where there are no contributions from line $e_1$. 
Hence the weight of $e_1$ 
has no impact on the matrix $\mathbf{X}_i \mathbf{X}_i^{\top}$ for $i = 1, ~ \cdots, ~ m-n+1$.  
Because the weights of the lines in $\mathcal{G}_2\cup\mathcal{G}_2$ are not changed, thus 
increasing the coupling strength of line $e_1$ in $\mathcal{G}_1$ has no impact on the variance 
of the phase difference in these lines.
 (iii) If $\mathcal{G}_1$ is a cycle-cluster, denote the number of lines in $\mathcal{G}_1$
 and $\mathcal{G}_2\cup\mathcal{G}_3$ by $N$ and $m-N$, the lines by $e_1,e_2,\cdots,e_N$ and 
 $e_{N+1}, ~ \cdots, ~ e_m$,  and the number of the fundamental cycles by $m_1$ and $m_2$ respectively.  
 Here $m_1+m_2=m-n+1$. 
Because the lines in one cycle-cluster are never in the other cycle-cluster, 
the basis vectors of the cycles in cycle-cluster $\mathcal{G}_1$ have the form 
$\mathbf{\xi}_i
  = \begin{bmatrix}
      \xi_{i,1} & \xi_{i,2} & \cdots & \xi_{i,N} & 0 & \cdots & 0
    \end{bmatrix}^{\top}
$
for $i=1,\cdots,m_1$ and 
those of the cycles in  $\mathcal{G}_2\cup\mathcal{G}_3$ have the form 
$\mathbf{\xi}_i=
  \begin{bmatrix}
    0 & 0 & \cdots & 0 & \xi_{i,{N+1}} & \cdots&\xi_{i,m}
  \end{bmatrix}^{\top}
$
for $i = m_1+1, ~ \cdots, ~ m-n+1$.
In these vectors, $\xi_{i,j}$ are either $1$,$-1$, or $0$. 
By the Gram-Schmidt orthogonalization of $\mathbf{R}^{-1/2} \mathbf{\xi}_i$, 
we obtain the orthonormal vectors
\begin{eqnarray*}
        \mathbf{X}_i
    & = & \begin{bmatrix}
            x_{i,1} & \cdots & x_{i,N} & 0 & \cdots & 0
          \end{bmatrix}^{\top}, ~
          \forall ~ i = 1, ~ \cdots, ~ m_1, \\
        \mathbf{X}_i
    & = & \begin{bmatrix}
            0 & \cdots & 0 & x_{i,{N+1}} & \cdots & x_{i,m}
          \end{bmatrix}^{\top}, ~
          \forall ~ i = m_1+1, ~ \cdots, ~ m-n+1.  
\end{eqnarray*} 
It is obvious that the entries in the first $N$ columns and 
the first $N$ rows of the matrix 
$\sum_{i=m_1+1}^{m-n+1} ~ \mathbf{X}_i \mathbf{X}_i^{\top}$ 
are all zero.  
This indicates that the lines in $\mathcal{G}_2\cup\mathcal{G}_3$
have no contributions to the first $N$ columns and 
the first $N$ rows of $\mathbf{Q}_y$.  
Similarly, the lines in cycle-cluster $\mathcal{G}_1$ have no contributions 
to the last $m-N$ columns and the last $m-N$ rows of $\mathbf{Q}_y$.
Hence, constructing new lines or increasing coupling strength of lines 
in cycle-cluster $\mathcal{G}_1$
has no impact on the variance of the phase differences 
in the lines in $\mathcal{G}_2\cup\mathcal{G}_3$ and vice versa.  
\par 
(iii) We first consider the case where the coupling strength
of a line in a cycle-cluster increases.  
If there are two or more cycle-clusters,
based on Corollary \ref{corollarymain} (ii), then it is deduced that increasing the coupling strength of a line in a cycle-cluster has no impact 
on the variances in the lines in the cycle-clusters that do not contain this line.  Thus, we focus only
on its impact on those in the cycle-cluster that contains this line. 
Assume there is only one cycle-cluster in the graph where the coupling strength of of line $e_1$ increases.  
Because the weight of line $e_1$ increases
 while those in the other lines are not changed, we only need to study the changes in the variances when the weight of line $e_1$ increase. The dimension $k$ of the kernel of $\mathbf{C} \mathbf{R}^{1/2}$ 
equals to $m-n+1$ and the basis vectors 
can be obtained from the $k$ fundamental cycles. 
The corresponding basis vectors to the $k-1$ cycles  
which do not include line $e_1$ 
have the form 
$\mathbf{\xi}_i=
  \begin{bmatrix}
    0 & \xi_{i,2} & \xi_{i,3} & \cdots & \xi_{i,m}
  \end{bmatrix}^{\top}
$
for $i = 1, ~ \cdots, ~ k-1$, where 
$\xi_{i,q} = 1, ~ -1, ~\text{or,} ~ 0$ for $q=2, ~ \cdots, ~ m$. 
The basis vector corresponding to the fundamental cycle 
which includes line $e_1$ has the form 
$\mathbf{\xi}_k=
  \begin{bmatrix}
    \xi_{k,1} & \xi_{k,2} & \cdots & \xi_{k,m}
  \end{bmatrix}^{\top}
$
where $\xi_{k,1} = 1, ~ \text{or,} ~ -1$ and 
$\xi_{k,q} = 1, ~ -1, ~ \text{or,} ~ 0$ 
for $q = 2, ~ \cdots, ~ m$. 
By the Gram-Schmidt orthogonalization of $\mathbf{R}^{-1/2} \mathbf{\xi}_i$,
we obtain
$\mathbf{X}_i=
  \begin{bmatrix}
    0 & x_{i,2} & x_{i,3} & \cdots & x_{i,m}
  \end{bmatrix}^{\top}
$
for $i = 1, ~ \cdots, ~ k-1$ 
where there are no contributions from line $e_1$.  Hence the weight of $e_1$ denoted by $l_1$ 
has no impact on the matrix 
$\mathbf{X}_i \mathbf{X}_i^{\top}$ for $i = 1, ~ \cdots, ~ k-1$. 
The last vector $\mathbf{X}_k$ can be obtained by normalization 
of the vector 
\begin{eqnarray*}
\mathbf{X}_k' = \mathbf{R}^{-1/2} \mathbf{\xi}_k
  - \alpha_1 \mathbf{X}_1
  - \cdots - \alpha_{k-1} \mathbf{X}_{k-1},
\end{eqnarray*}
where $\alpha_i=\frac{(\mathbf{R}^{-1/2}\mathbf{\xi}_k)^\top\mathbf{X}_i}{\mathbf{X}_i^\top\mathbf{X}_i} \in \mathbb{R}$ is independent 
of $l_1$ because the first element of $\mathbf{X}_i$ equals zero for $i = 1, ~ \cdots, ~ k-1$.
Hence $\mathbf{X}_k'$ has the from 
\begin{eqnarray*}
\mathbf{X}_k'=
  \begin{bmatrix}
    l_1^{-1/2}\xi_{k,1} & x'_{k,2} & x'_{k,3} & \cdots & x'_{k,m}
  \end{bmatrix}^{\top}
\end{eqnarray*}
where $x'_{k,q}$ is independent of $l_1$ for $q = 2, ~ \cdots, ~ m$. 
By the normalization of $\mathbf{X}_k'$,
we obtain 
$\mathbf{X}_k = a \mathbf{X}_k'$ where 
$a = 
  \big(
    l_1^{-1} + \sum_{i=2}^{m} ~ {x'_{k,i}}^2
  \big)^{-1/2}$. 
Hence, 
the diagonal element of $\mathbf{X}_k \mathbf{X}_k^{\top}$ equals to   
$a^2l_1^{-1}$ for $i=1$ and $a^2{x'_{k,i}}^2$ for $i = 2, ~ \cdots, ~ m$. 
Thus,
the variance of the phase difference in line $e_1$ equals to 
$\frac{1}{2}\beta(l_1^{-1}-a^2l_1^{-2})$ and 
that in line $e_q$ with weight $l_q$ equals to 
$\frac{1}{2} \beta l_q^{-1} 
  (1-a^2{x'_{k,q}}^2 
    - \sum_{i=1}^{k-1} ~{x_{i,q}^2}
  )$ 
for $q=2,\cdots,m$. 
Clearly, the variance in line $e_1$ decreases as $l_1$ increases. Here we further prove this trend also holds for the variance 
in line $e_1$. Let $z=l_1^{-1}$ and $b=\sum_{i=2}^{m} ~ {x'_{k,i}}^2,$ then the variance in line $e_1$ becomes
\begin{align*}
\frac{1}{2}\beta(l_1^{-1}-a^2l_1^{-2})=\frac{1}{2}\beta(z-(z+b)^{-1}z^2).
\end{align*}
The derivative of this variance with respect to $z$ is $\frac{b^2\beta}{2(z+b)^2}$, which 
is strictly positive for any $z>0$. Hence, the variance is monotonously increasing with respect to $z$ and decreases 
as $l_1$ increases. 
\par
We next consider the case when a new line is constructed 
in a cycle-cluster without changing the weight of all the other lines. 
Assume line $e_1$ is the new line. 
The variance of the phase difference in line $e_q$ with weight $l_q$  
before adding line $e_1$ is 
\begin{eqnarray*}
\frac{1}{2}\beta l_q^{-1}(1-\sum_{i=1}^{k-1}{x_{i,q}^2})
\end{eqnarray*}
for $q = 2, ~ \cdots, ~ m$, which decreases to 
\begin{eqnarray*}
\frac{1}{2}\beta l_q^{-1}
  (
    1-a^2{x'_{k,q}}^2
    - \sum_{i=1}^{k-1}{x_{i,q}^2}
  )
\end{eqnarray*}
after adding line $e_1$. 
\par 
(iv) Denote the vector corresponding 
to this cycle by $\mathbf{\xi}_1$ and the ones corresponding to 
the other cycles by $\mathbf{\xi}_i$ with $i = 2, ~ \cdots, ~ m-n+1$.
Without loss of generality, 
we assume the lines $e_1,e_2,\cdots,e_N$ 
with weights $l_1, ~ l_2, ~ \cdots, ~ l_N$ 
are in the cycle and 
the direction of these lines are consistent 
with the direction of the cycle. 
By the definition of the basis of the cycle space, 
we obtain 
$\mathbf{\xi}_1=
  \begin{bmatrix}
    1 & 1 & \cdots & 1 & 0 & \cdots & 0
  \end{bmatrix}^{\top}
$ 
where the first $N$ elements equal to 1 and the last $m-N$ elements 
equal to 0, and 
$\mathbf{\xi}_i=
  \begin{bmatrix}
    0 & 0 & \cdots & 0 & \xi_{i,{N+1}} & \cdots & \xi_{i,m}
  \end{bmatrix}^{\top}
$
where the first $N$ elements
are all 0 and the last $m-N$ elements equal to either 0 or 1. 
Obviously, the vector $\mathbf{R}^{-1/2} \mathbf{\xi}_1$ is orthogonal 
to the vector $\mathbf{R}^{-1/2} \mathbf{\xi}_i$ 
for $i = 2, ~ \cdots, ~ m-n+1$.  
By scaling these vectors to unit Euclidian length,  
we obtain the unit vector 
\begin{eqnarray*}
        \mathbf{X}_1
    & = & \Big(
            \sum_{k=1}^N ~ l_{k}^{-1}
          \Big)^{-1/2}
          \begin{bmatrix}
            l_1^{-1/2} & l_2^{-1/2} & \cdots & l_N^{-1/2} & 0 & \cdots & 0
          \end{bmatrix}^{\top},\mbox{~~for} ~ \mathbf{R}^{-1/2} \mathbf{\xi}_1,
\end{eqnarray*}
and
$\mathbf{X}_i=
  \begin{bmatrix}
    0 & 0 & \cdots & 0 & x_{i,{N+1}} & \cdots & x_{i,m}
  \end{bmatrix}^{\top}
$ for the linear subspace composed by the vectors 
$\mathbf{R}^{-1/2} \mathbf{\xi}_i$ 
for $i = 2, ~ \cdots, ~ m-n+1$.  
Since the first $N$ elements are all $0$,  
$\mathbf{X}_i \mathbf{X}_i^{\top}$ for $i = 2, ~ \cdots, ~ m-n+1$  
has no contributions to the first $N$ columns and
the first $N$ rows of $\mathbf{Q}_y$. 
By (\ref{theoremmain-1}), 
we obtain that the $k$th diagonal element of 
$\mathbf{Q}_y$ for $k = 1, ~ \cdots, ~ N$ is
\begin{eqnarray*}
\frac{\beta}{2}
  \Big(
    l_{k}^{-1}-l_{k}^{-2}
    \big(
      \sum_{r=1}^N ~
        l_{r}^{-1}
    \big)^{-1}
  \Big)
\end{eqnarray*}
from which we obtain (\ref{eq:singlecycletwo}) 
by replacing $l_k$ by $w_{ij}$. 
If $l_k=\gamma$ for all $k=1,\cdots,N$, 
we further get the first $N$ diagonal elements 
of $\mathbf{Q}_y$ equal to 
$\frac{\beta}{2\gamma}(1-\frac{1}{N})$.
\end{proof}
\begin{proof}
Proof of Theorem \ref{matrixbounds}.
\par 
With the definition of $\underline{\beta}$ and $\overline{\beta}$,  
we define $\underline{\mathbf B}=(\underline{\beta}\mathbf D)^{1/2}$ and 
$\overline{\mathbf B}=(\overline{\beta}\mathbf D)^{1/2}$. Thus, 
\begin{eqnarray*}
\underline{\beta}\mathbf{I}_n=\mathbf{D}^{-1/2} \underline{\mathbf{B}}\underline{\mathbf{B}}^\top\mathbf{D}^{-1/2}
  \preceq \mathbf{D}^{-1/2} \mathbf{B}\mathbf{B}^\top\mathbf{D}^{-1/2}
  \preceq \mathbf{D}^{-1/2} \overline{\mathbf{B}}\overline{\mathbf{B}}^\top\mathbf{D}^{-1/2}
  =\overline{\beta}\mathbf{I}_n.
\end{eqnarray*}
From formula (\ref{LyapunovQnplus1}), 
we derive 
$\frac{1}{2} ~ \underline{\beta}
  \mathbf{\Lambda}_{n-1}^{-1}
  \preceq
  \mathbf{Q}_{n-1}
  \preceq
  \frac{1}{2} \overline{\beta} \mathbf{\Lambda}_{n-1}^{-1}
$. 
Thus, it yields by (\ref{Qdelta}) that 
\begin{eqnarray*}
    &   & \frac{1}{2}
          \underline{\beta}
          \mathbf{C}^{\top}
          \mathbf{D}^{-1/2}
          \mathbf{U}_2
          \mathbf{\Lambda}_{n-1}
          \mathbf{U}_2^{\top}
          \mathbf{D}^{-1/2}
          \mathbf{C}
          \preceq
            \mathbf{Q}_{\mathbf{y}}
          \preceq
            \frac{1}{2}
            \overline{\beta}
            \mathbf{C}^{\top}
            \mathbf{D}^{-1/2}
            \mathbf{U}_2
            \mathbf{\Lambda}_{n-1}
            \mathbf{U}_2^{\top}
            \mathbf{D}^{-1/2} 
            \mathbf{C}.
\end{eqnarray*}
By repeating the proof of Theorem \ref{theoremmain}, 
we obtain the bounds of $\mathbf{Q}_{\mathbf{y}}$ in (\ref{bounds}). 
\end{proof}
\section{Conclusions}
\label{sec:conclusions}
Explicit expressions for the asymptotic variance matrix
of a stochastic linear system for the evaluation of the fluctuations in complex phase oscillators
have been derived and analyzed.
\par
Research interest remains for optimization of the network topology to improve the synchronization stability of the complex 
phase oscillators. 
%
\bibliography{sample,base,basetmp}
\bibliographystyle{siamplain}
\end{document}